\documentclass[%
 reprint,
 superscriptaddress,preprintnumbers,
 nofootinbib,
 amssymb,
 aps,
]{revtex4-1}

\usepackage[utf8]{inputenc}
\usepackage{hyperref}
\usepackage{amsmath,amssymb,mathtools}
\usepackage{dsfont}
\usepackage{graphicx}   
\usepackage[table]{xcolor}
\usepackage{colortbl}
\usepackage{url}
\usepackage{slashed}
\usepackage[noindentafter]{titlesec} 
\allowdisplaybreaks

\usepackage{siunitx}
\usepackage{xspace}

    \makeatletter
    \def\CT@@do@color{%
      \global\let\CT@do@color\relax
            \@tempdima\wd\z@
            \advance\@tempdima\@tempdimb
            \advance\@tempdima\@tempdimc
    \advance\@tempdimb\tabcolsep
    \advance\@tempdimc\tabcolsep
    \advance\@tempdima2\tabcolsep
            \kern-\@tempdimb
            \leaders\vrule
                    \hskip\@tempdima\@plus  1fill
            \kern-\@tempdimc
            \hskip-\wd\z@ \@plus -1fill }
    \makeatother

\usepackage{accents}
\DeclareMathSymbol{\widetildesym}{\mathord}{largesymbols}{"65}
    \newcommand\lowerwidetildesym{%
      \text{\smash{\raisebox{-1.3ex}{%
        $\widetildesym$}}}}

\newcommand\parwidetilde[1]{%
   \mathchoice
      {\accentset{\displaystyle\scalebox{.3}{(}\lowerwidetildesym\scalebox{.3}{)}}{#1}}
         {\accentset{\textstyle\scalebox{.3}{(}\lowerwidetildesym\scalebox{.3}{)}}{#1}}
         {\accentset{\scriptstyle\scalebox{.3}{(}\lowerwidetildesym\scalebox{.3}{)}}{#1}}
         {\accentset{\scriptscriptstyle\scalebox{.3}{(}\lowerwidetildesym\scalebox{.3}{)}}{#1}}
   }

\def\thesubsection{\arabic{section}.\arabic{subsection}}

\def\thesection{\arabic{section}}
\titleformat*{\subsubsection}{\normalfont \small \bfseries \boldmath}
\renewcommand{\paragraph}[1]{\vspace{.3em} \indent {\bfseries \boldmath #1 ---}\xspace }
\makeatletter
    \renewcommand{\p@subsection}{}
    \renewcommand{\p@subsubsection}{}
\makeatother

\definecolor{red}{rgb}{0.6,.0706,.1373}
\definecolor{blue}{rgb}{0,0.396,0.741}
\newcommand\myshade{80}
\colorlet{mylinkcolor}{violet}
\colorlet{mycitecolor}{violet}
\colorlet{myurlcolor}{violet}

\hypersetup{
  linkcolor  = mylinkcolor!\myshade!black,
  citecolor  = mycitecolor!\myshade!black,
  urlcolor   = myurlcolor!\myshade!black,
  colorlinks = true
}
\newcommand{\U}{\mathrm{U}}
\newcommand{\SU}{\mathrm{SU}}

\newcommand{\LL}{\mathrm{L}}
\newcommand{\RR}{\mathrm{R}}
\newcommand{\ud}[2]{\phantom{}^{#1}\phantom{}_{#2}}
 
\newcommand{\diag}{\mathop{\mathrm{diag}}}
\renewcommand{\Im}{\mathop{\mathrm{Im}}}
\renewcommand{\Re}{\mathop{\mathrm{Re}}}
\newcommand{\eminus}{\vcenter{\hbox{\scalebox{0.6}[1]{$ - $}}}}	
\newcommand{\rep}[1]{\mathbf{#1}}

\newcommand{\sscript}[1]{{\scriptscriptstyle \mathrm{#1}}}

\def\L{\mathcal{L}}
\newcommand{\gf}{$ \SU(2)_{q+\ell}$\xspace}


\begin{document}


\title{
Rising Through the Ranks: Flavor Hierarchies from a Gauged \boldmath $ \mathrm{ SU(2)}$ Symmetry
}

\author{Admir Greljo}
\email{admir.greljo@unibas.ch}
\author{Anders Eller Thomsen}
\email{thomsen@itp.unibe.ch}
\affiliation{Department of Physics, University of Basel, Klingelbergstrasse 82,  CH-4056 Basel, Switzerland}



\begin{abstract}

We propose an economical model to address the mass hierarchies of quarks and charged leptons. 
The light generations of the left-handed fermions form doublets under an $ \SU(2) $ flavor symmetry, which is gauged. The generational hierarchies emerge from three independent rank-one contributions to the Yukawa matrices: one is a renormalizable contribution, the second is suppressed by a mass ratio, and the last by an additional loop factor. The model is renormalizable and features only a handful of new fields and is remarkably simple compared to typical completions of gauged flavor symmetries or Froggatt-Nielsen. The model has a rich phenomenology, and we highlight promising signatures, especially in the context of $K$ and $B$ meson physics. This includes an interpretation of the latest $B^+ \to K^+ \nu \bar \nu$ measurement from Belle II. 

\end{abstract}

\maketitle

\section{Introduction} 
\label{sec:intro}

The flavor hierarchies observed in particle physics stand as a captivating enigma. In the Standard Model (SM), fermion masses and mixings originate from the interactions with a single Higgs field, $H$. Three Yukawa couplings (square matrices of dimension three) are associated with up- and down-type quarks and charged leptons. The eigenvalues of these matrices dictate the spectrum of charged fermion masses, which are intrinsically arbitrary parameters of the theory. 

We observe a consistent hierarchy spanning 1--2 orders of magnitude between Higgs couplings to consecutive generations of fermions, consistently through all three generations. They range from $y_e \sim 10^{\eminus 6}$ for the electron to $y_t \sim 1$ for the top quark. 
Also the misalignment between up and down Yukawa matrices captured by the CKM mixing matrix ($V_{ij}$) reveals a hierarchical pattern. It is characterized by $V_{ii}\simeq 1$ ($i=1,2,3$) and $V_{ii} \gg V_{12} \gg V_{23} \gg V_{13}$, indicating only modest inter-family mixings. 
While the Yukawa parameters are technically natural according to 't Hooft~\cite{tHooft:1979rat}, it is hard to accept such a hierarchical pattern within the SM, a theory where the relevant input parameters all enter at the same footing. The puzzle is likely to be resolved with physics beyond the SM.

Addressing the enigma of flavor hierarchies prompts the question: what truly constitutes a solution within a quantum field theory? Often, new models bring forth a plethora of new parameters, transforming the inquiry into a question of statistical distribution. Our guiding principle in this paper is the introduction of new marginal interactions characterized by parameters around $\mathcal{O}(0.3)$ or within the range $[0.1, 1]$ (avoiding Landau poles when extrapolating to high energies typically requires values $\lesssim 1$). Yet, when many parameters are present, a few from the tail of the distribution might appear deceptively small. The balancing act lies in the complexity of these models versus the broadness of the parameter distributions.

A review of prior attempts to address flavor hierarchies brings several solutions to the fore. A non-exhaustive list of representative examples include Froggatt-Nielsen models~\cite{Froggatt:1978nt, Leurer:1992wg, Leurer:1993gy, Fedele:2020fvh, Cornella:2023zme, Asadi:2023ucx, Smolkovic:2019jow}, gauged flavor symmetries~\cite{Grinstein:2010ve, DAgnolo:2012ulg, Kaplan:1993ej, King:2003rf, Antusch:2008jf, Antusch:2007re, Bishara:2015mha, Linster:2018avp, Davighi:2022fer}, accidental flavor~\cite{Barr:1990td, Barr:2001vj, Ferretti:2006df}, radiative mass models~\cite{Weinberg:1972ws, Arkani-Hamed:1996kxn, Altmannshofer:2014qha, Weinberg:2020zba, Baker:2020vkh}, multi-Higgs models~\cite{Porto:2007ed,Hill:2019ldq,Baek:2023cfy}, multiscale flavor~\cite{Barbieri:1994cx, Panico:2016ull, Bordone:2017bld, Greljo:2018tuh, Allanach:2018lvl, Greljo:2019xan,Allwicher:2020esa, Davighi:2023iks, Davighi:2023evx, FernandezNavarro:2023rhv}, warped compactification~\cite{Randall:1999ee, Arkani-Hamed:1999ylh, Fuentes-Martin:2022xnb}, partial compositeness~\cite{Kaplan:1991dc, Ferretti:2014qta, Sannino:2016sfx}, clockwork~\cite{Patel:2017pct, Alonso:2018bcg, Altmannshofer:2021qwx}, modular symmetries~\cite{Feruglio:2021dte}, etc. An underlying theme amongst many of these mechanisms is the complexity of their typical ultraviolet (UV) completions. For instance, Froggatt-Nielsen often demands intricate chains of vector-like fermions~\cite{Smolkovic:2019jow}, while gauged flavor symmetries typically require complicated spontaneous symmetry-breaking sectors~\cite{Alonso:2011yg}. 

In this paper, we propose a streamlined model that integrates features from various mechanisms previously mentioned. While common gauged flavor models often start with large subgroups of $\U(3)^5$ global group~\cite{DAmbrosio:2002vsn} (or $\U(2)^5$~\cite{Barbieri:1995uv, Barbieri:2011ci, Kagan:2009bn}), we consider a rather small subgroup. The $\SU(2)_{q}$ global symmetry of the left-handed quarks emerges as an approximate symmetry of the quark sector, as was nicely illustrated in the recent work~\cite{Bento:2023owf} on SM flavor invariants. The analysis highlights that SM parameters offer no insights regarding the right-handed symmetry groups.

Building on this insight, the keystone of our model is the gauged \gf flavor (horizontal) symmetry group, under which light generations of left-handed quarks and leptons transform as doublets~\cite{Babu:1990hu, Shaw:1992gk, Darme:2023nsy}. This choice stands out as it ensures an anomaly-free gauge symmetry without introducing additional chiral fermions. Moreover, the mass hierarchies of charged leptons mirror those observed in quarks. The symmetry selection rules implied at the level of renormalizable interactions predict that the three Yukawa matrices are all of rank one, resulting in one massive and two massless generations. This provides a basis for coherently explaining the observed hierarchy between the third and light generations.

The lightness of the first two families is attributed to the gauge-invariant dimension-5 operators (e.g. $\bar q H \Phi d$) built with a symmetry-breaking scalar doublet $\Phi$, with a VEV structure $\langle \Phi \rangle \gg \langle H \rangle$. Interestingly, when one attempts to generate this effective interaction at tree level with a minimal set of new heavy vector-like fermions (VLF) denoted as $F$ with $M_F \gg \langle \Phi \rangle$, the resulting Yukawa matrices become of rank two, predicting the first generation to be massless while giving a small mass to the second generation. This sets the basis for the hierarchy between the first two generations.

\begin{figure}[t]
         \centering
         \includegraphics[width=0.49\textwidth]{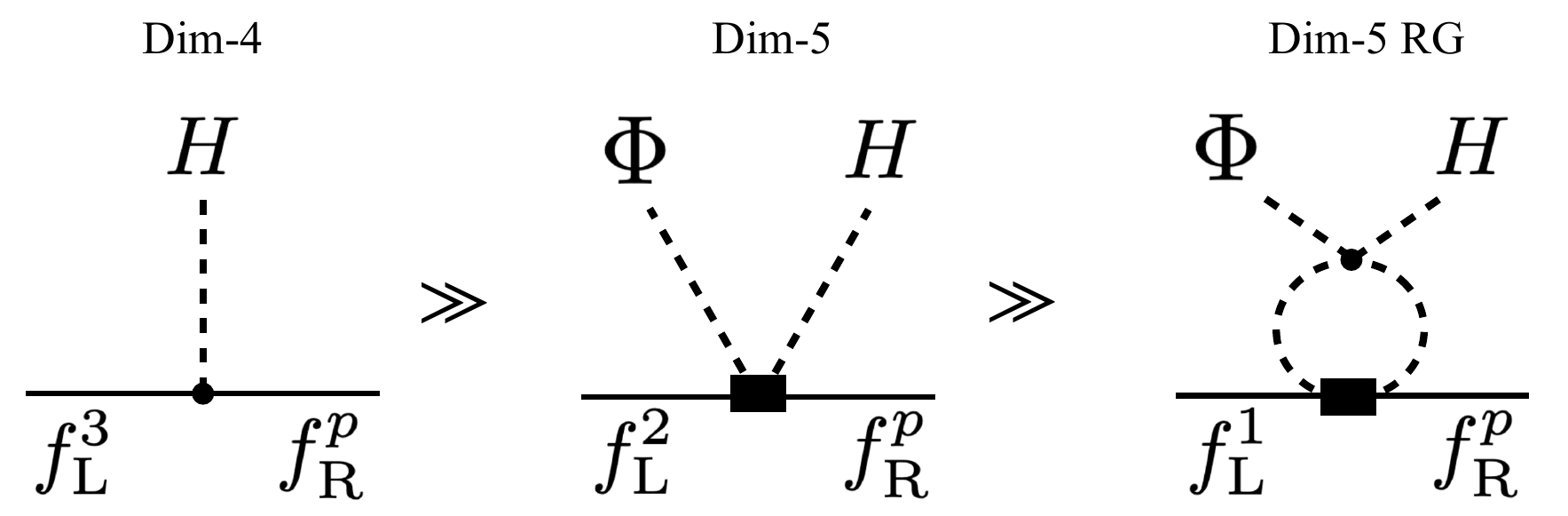}
        \caption{The mechanism for flavor hierarchies. The leading gauge invariant operators under \gf horizontal gauged symmetry matching to the SM after SSB by $\langle \Phi \rangle \gg \langle H \rangle$.}
        \label{fig:mechanism}
\end{figure}

\begin{figure}
         \centering
         \includegraphics[width=0.25\textwidth]{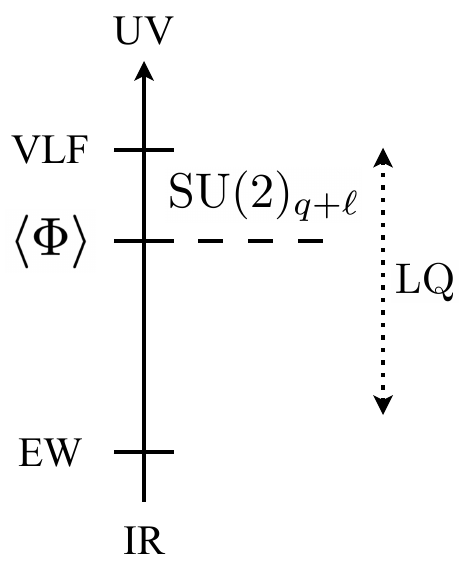}
        \caption{The schematic spectrum of the UV completion for operators in Fig.~\ref{fig:mechanism}. }
        \label{fig:scales}
\end{figure}

Finally, to generate hierarchically suppressed masses for the first generation, we opt for generating linearly independent contributions to the dimension-5 operators at the loop level. 
Instead of introducing new fields above the UV scale $M_F$, we introduce new IR states, scalar leptoquarks (LQ) whose masses can be anywhere below the UV scale and which play a role in renormalization group (RG)--induced dimension-5 operators, see Figs.~\ref{fig:mechanism} and \ref{fig:scales}. The emerging SM flavor structure is largely insensitive to the masses of these scalars as long as they are light compared to the UV scale. This provides a robust structure in which the hierarchies are controlled solely by the ratio $\langle \Phi \rangle / M_F$. 

This paper is organized as follows: Section~\ref{sec:model} presents our model, anchoring our discussion on the rank of Yukawa matrices and detailing how the model generates the SM flavor patterns. In Section~\ref{sec:pheno}, we delve into the rich phenomenology of the model, focusing on the flavored gauge bosons, vector-like fermions, and leptoquarks. In conclusion, Section~\ref{sec:conc} offers our perspective on the remaining open questions of the model, weaving this mechanism into the Pati-Salam quark-lepton unification framework.

\section{Model} 
\label{sec:model}

This section constructs our model for flavor hierarchies. It discusses in detail the three steps required to fill the rank of the Higgs Yukawa coupling matrices. 

\subsection{Rank 1} 
\label{sec:rank1}
The basis of our model is a version of the SM where the two light generations of left-handed fermions form doublets under a new gauge symmetry \gf. The charges of the fields are detailed in Table~\ref{tab:field_content_SM}. We anticipate having to break the new symmetry at a high energy scale and introduce a new SM-singlet scalar doublet $ \Phi $ of \gf, whose VEV will do the trick. The choice to use a doublet is determined because it allows for constructing dimension-5 mass operators for the first two generations of SM fermions. Without further exploring the scalar potential, we take the VEV of the doublet to be
    \begin{equation}
    \langle \Phi^\alpha \rangle = \binom{0}{v_\Phi }~,
    \end{equation}
which we treat as a free scale of the model. $ \alpha, \beta,\ldots $ denote doublet indices of \gf.  

As indicated, the \gf gauge symmetry prevents marginal Higgs Yukawa couplings with the light generations of the left-handed fermions. The field content in Table~\ref{tab:field_content_SM} allows only for the Yukawa interactions\footnote{We keep $SU(2)_\LL $  and $ \SU(3)_c $ indices implicit throughout the paper to avoid unnecessary clutter.}
    \begin{equation} \label{eq:3rd_gen_Yukawas}
    \L \supset  -x_u^{p} \, \overline{q}^3 \widetilde{H} u^p -x_d^{p} \, \overline{q}^3 H d^p -x_e^{p} \, \overline{\ell}^3 H e^p +\text{H.c.}~,
    \end{equation}
where $ \widetilde{H}^i = \varepsilon^{ij} H^\ast_j $ (and $ \widetilde{\Phi}^\alpha = \varepsilon^{\alpha \beta} \Phi^\ast_\beta $).\footnote{We employ the conventions that the antisymmetric tensor of $ \SU(2) $ has $ \varepsilon^{12} = -\varepsilon^{21} = 1 $. For later convenience, we define $ \widetilde{X}^i = \varepsilon^{ij} X^\ast_j $ for any $ \SU(2)_L $ doublet field $X$.}
The Yukawa couplings $ x_f^p $ are all three-dimensional vectors with flavor indices $ p,r,\ldots  \in \{1,2,3\} $. 

At this stage, no other independent Yukawa couplings can be generated. The  $ 3\times 3$ SM Yukawa coupling matrices, $ Y_f^{pr} $ resulting from this theory at low energies will necessarily have rank 1; each is constructed with just one vector $ x_f^r $. For practical purposes, the flavor symmetry of the right-handed quarks can be used to remove unphysical parameters from the couplings, which can be parametrized by
    \begin{equation} \label{eq:coupling_parametrization_1}
    x_f^p = \big(0,\, 0,\, x_{f3} \big), \qquad x_{f3} \in \mathbb{R}_0^+ ,
    \end{equation}
for $ f \in \{ u,\, d, \,e\}$.
The \gf symmetry establishes a hierarchy for the Yukawa couplings between the third and the light generations. While having massless first and second-generation fermions is unphysical, masses can be engineered as perturbations on our foundation.

\begin{table} 
	\begin{center} \renewcommand{\arraystretch}{1.1}
	\begin{tabular}{|c|c|c|c|c|}
	\hline  \rowcolor{black!15}
	Field & $ \SU(3)_c $ & $ \SU(2)_\LL $ & $ \U(1)_Y $ & $ \SU(2)_{q+\ell}$ \\
	\hline
	$ q^\alpha_\LL $ & $ \rep{3} $ & $ \rep{2} $ & $ 1/6 $ & $ \rep{2} $ \\ 
	$ q^3_\LL $ & $ \rep{3} $ & $ \rep{2} $ & $ 1/6 $ & $ \rep{1} $ \\ 
	$ u^p_\RR $ & $ \rep{3} $ & $ \rep{1} $ & $ 2/3 $ & $ \rep{1} $ \\ 
	$ d^p_\RR $ & $ \rep{3} $ & $ \rep{1} $ & $ \eminus1/3 $ & $ \rep{1} $ \\ \hline 
	$ \ell^\alpha_\LL $ & $ \rep{1} $ & $ \rep{2} $ & $ \eminus 1/2 $ & $ \rep{2} $ \\ 
	$ \ell^3_\LL $ & $ \rep{1} $ & $ \rep{2} $ & $ \eminus 1/2 $ & $ \rep{1} $ \\ 
	$ e^p_\RR $ & $ \rep{1} $ & $ \rep{1} $ & $ \eminus 1 $ & $ \rep{1} $ \\ \hline
	$ H $ & $ \rep{1} $ & $ \rep{2} $ & $ 1/2 $ & $ \rep{1} $ \\ \hline \hline
	$ \Phi $ & $ \rep{1} $ & $ \rep{1} $ & $ 0 $ & $ \rep{2} $ \\ \hline
	\end{tabular}
	\caption{The SM gauge symmetry is extended with an \gf of the light left-handed families ($\alpha=1,2$). The theory is supplemented with the complex scalar field $ \Phi $, responsible for breaking \gf. This setup produces Yukawa matrices of rank~1, setting the basis for the hierarchy between the first two and the third generation. }
	\label{tab:field_content_SM}
	\end{center}
\end{table}

\subsection{Rank 2} 
\label{sec:rank2}
The next step towards a realistic IR flavor structure is a mechanism to generate higher dimensional Higgs--Yukawa operators involving the symmetry-breaking scalar $ \Phi $. Since $ \widetilde{\Phi} $ and $ \Phi $ share quantum numbers, it is possible to construct a pair of dimension-5 operators, such as $ \Phi^\alpha \bar q_\alpha H d^p $ and $ \widetilde{\Phi}^\alpha \bar q_\alpha H d^p $, for each fermion species. If both of these are introduced with uncorrelated couplings of the same size, there will be no splitting between first- and second-generation masses. This is a challenge to many potential UV completion of the dimension-5 operators. For instance, a completion through VLF partners of the right-handed SM fermions with \gf charges fails to explain the $1 - 2$ hierarchy without resorting to some ad hoc symmetry. 

\begin{figure*}
    \centering \includegraphics[]{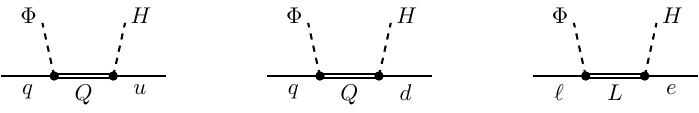}
    \caption{Feynman diagrams depicting mass generation for the \textit{second} SM family.}
    \label{fig:tree-level_masses}
\end{figure*}

A minimal mechanism for completing the dimension-5 operators includes vector-like fermions $ Q $ and $ L $ charged as the third-generation left-handed SM quarks and leptons, respectively. Their gauge charges are given in Table~\ref{tab:field_content_VLFs}. The new Yukawa couplings are
\begin{equation}\label{eq:4}
    \begin{split}
    \L \supset \,& + \big(y_q \Phi^\alpha + \tilde{y}_q \widetilde{\Phi}^\alpha\big) \overline{q}_\alpha Q + \big(y_\ell \Phi^\alpha + \tilde{y}_\ell \widetilde{\Phi}^\alpha\big) \overline{\ell}_\alpha L \\
    &- y_u^p \,\overline{Q} \widetilde{H} u^p - y_d^p \,\overline{Q} H d^p - y_e^p \,\overline{L} H e^p + \text{H.c.}~.
    \end{split}
\end{equation}
Integrating out $ Q $ and $ L $ at the tree level (see Fig.~\ref{fig:tree-level_masses}) results in the effective dimension-5 operators giving masses to light generation fermions. Furthermore, the limit $ M_Q, M_L \gg v_\Phi $ explains the hierarchy between the third and light generations. 

\begin{table}[b]
    \begin{center} \renewcommand{\arraystretch}{1.1}
    \begin{tabular}{|c|c|c|c|c|}
    \hline  \rowcolor{black!15}
    Field & $ \SU(3)_c $ & $ \SU(2)_\LL $ & $ \U(1)_Y $ & $ \SU(2)_{q+\ell}$ \\
    \hline
    $ Q_{\LL,\RR} $ & $ \rep{3} $ & $ \rep{2} $ & $ 1/6 $ & $ \rep{1} $ \\ 
    $ L_{\LL,\RR} $ & $ \rep{1} $ & $ \rep{2} $ & $ \eminus 1/2 $ & $ \rep{1} $ \\ \hline 
    \end{tabular}
    \caption{Vector-like fermion representations $ Q $ and $ L $. Their presence lifts the rank of the Yukawa matrices to rank~2, providing masses to the second generation.}
    \label{tab:field_content_VLFs}
    \end{center}
\end{table}

As advertised, all new contributions to the SM Yukawa matrices $ Y_f^{pr} $ from the dimension-5 operators are proportional to the new vector couplings $ y_f^r $. Thus, VLFs are sufficient only to raise the rank of the matrices from one to two. This construction, therefore, leaves the first-generation fermions massless while it explains the hierarchy between the massive second and third generations. 

The $ \Phi $ kinetic term exhibits an $ \SU(2) $ flavor symmetry between $ \Phi $ and $ \widetilde{\Phi} $, which share quantum numbers. This enhanced symmetry of the doublet is reminiscent of the approximate custodial symmetry of the Higgs field in the SM. Along with the remaining flavor symmetries after introducing the $ x^p_f $ couplings, it is used to eliminate redundant, unphysical parameters from the couplings. We can choose to parameterize the physical VLF Yukawa couplings as 
\begin{equation} \label{eq:coupling_parametrization_2}
	\begin{gathered}
	y_f^p = \big(0,\, y_{f2},\, y_{f3} \big), \qquad \tilde{y}_q =0\,, \\ y_{f2},\, y_{d3},\, y_{e3},\, y_q,\, y_\ell,\, \tilde{y}_\ell \in \mathbb{R}_0^+,\qquad  y_{u3} \in\mathbb{C}~.
	\end{gathered}
\end{equation}
This choice simplifies expressions in the following analysis.

\subsection{Rank 3} 
\label{sec:rank3}
A simple way to provide masses to the first generation is to introduce a second generation of VLF, whose couplings to the right-handed fermions would lift the rank of $ Y_f^{pr} $ to three. However, barring conspiracies between couplings, a hierarchy between the VLF masses would be required to account for the hierarchy between first and second-generation SM masses. 
We propose an alternative scenario where the first-generation Yukawa couplings are suppressed by a loop factor compared to the second-generation. This is a more robust approach, and it does not introduce a new heavy scale to the problem. 

A minimal loop completion of the dimension-5 operators uses the VLF $ Q $ and $ L $ already present in the model. This identifies the scale of the loops with that of the tree-level contribution. To lift the rank of the Higgs-Yukawa couplings, one must introduce new vector couplings (in flavor space), which can come from coupling $ L $ to $ u/d $ and $ Q $ to $ e $ via LQ interactions. To make a long story short, we introduce three scalar LQ fields $ R_u $, $ R_d $, and $ S $ with charges listed in Table~\ref{tab:field_content_LQs} to complete the picture.

\begin{table}[b]
	\begin{center} \renewcommand{\arraystretch}{1.1}
	\begin{tabular}{|c|c|c|c|c|}
	\hline  \rowcolor{black!15}
	Field & $ \SU(3)_c $ & $ \SU(2)_\LL $ & $ \U(1)_Y $ & $ \SU(2)_{q+\ell}$ \\
	 \hline 
	$ R_u $ & $ \rep{3} $ & $ \rep{2} $ & $ 7/6 $ & $ \rep{1} $ \\ 
	$ R_{d} $ & $ \rep{3} $ & $ \rep{2} $ & $ 1/6 $ & $ \rep{1} $ \\ 
	$ S  $ & $ \rep{3} $ & $ \rep{1} $ & $ 2/3 $ & $ \rep{2} $ \\ \hline 
	\end{tabular}
	\caption{Scalar leptoquark fields contribute to radiative mass generation in the first family, resulting in rank-3 Yukawa matrices.}
	\label{tab:field_content_LQs}
	\end{center}
\end{table}

\begin{figure*}
    \centering \includegraphics[]{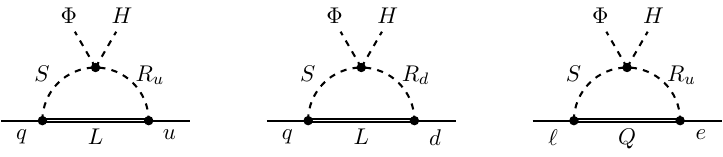}
    \caption{Feynman diagrams depicting mass generation for the \textit{first} SM family.}
    \label{fig:loop-level_masses}
\end{figure*}

The LQ fields have the Yukawa interactions  
    \begin{equation}\label{eq:lag_LQ_VLF}
    \begin{split}
    \L \supset\,&  - z^p_u\, \overline{L} u^p \widetilde{R}_u - z^p_d\,\overline{L} d^p \widetilde{R}_{d} - z^p_e\, \overline{Q} e^p R_u \\
    & - z_q\, \overline{q}_\alpha L S^{ \alpha} - z_\ell\, \overline{\ell}_\alpha Q \widetilde{S}^{\alpha} + \text{H.c.}~,
    \end{split}
    \end{equation}
involving the VLF. With the LQs added to the model, there are new loop contributions to the mass-generating dimension-5 operators as shown in Fig.~\ref{fig:loop-level_masses}. The mechanism also relies on a set of quartic couplings,
    \begin{multline}\label{eq:crossquartics}
    V \supset  (\lambda_{u} \Phi^\alpha+ \tilde{\lambda}_{u} \widetilde{\Phi}^\alpha) S_{\alpha}^\ast R_u H^\ast  \\ 
    + (\lambda_{d} \Phi^\alpha + \tilde{\lambda}_{ d} \widetilde{\Phi}^\alpha)  S_{\alpha}^\ast R_{d} \widetilde{H}^\ast + \text{H.c.}~,
    \end{multline}
closing the LQ loops. The new vector couplings $ z_f^p $ are sufficient to lift the effective Yukawa coupling matrices of the SM fermions to rank 3, thereby giving masses to the first family. The additional loop suppression relative to the tree-level Yukawas of the second generation accounts for the 1--2 hierarchy. 

The phenomenology of the LQs is mostly dictated by their couplings to pairs of SM fermions:
    \begin{equation} \label{eq:lag_LQ_SM_fermions}
    \L \supset - \kappa_{ u}^p\, \overline{\ell}^3 \widetilde{R}_u u^p - \kappa_{d}^p\, \overline{\ell}^3 \widetilde{R}_{d} d^p - \kappa_{e}^p\, \overline{q}^3 R_u e^p + \text{H.c.}\,.
    \end{equation}
The \gf symmetry dictates a particular structure to these couplings as compared to vanilla LQs. All such couplings of $ S $ are prohibited whereas the remaining couplings of $ R_u $ and $ R_d $ couple exclusively to third-generation left-handed fermions at the level of marginal couplings. The result is a suppressed coupling to light left-handed mass eigenstates. Naively, one expects anarchic couplings between all three generations of right-handed fermions. 

There are a few unphysical phases in the model, which allow us to parametrize the new LQ parameters by
	\begin{equation} \label{eq:coupling_parametrization_3}
	\begin{gathered}
	z_f^p = \big(z_{f1},\, z_{f2},\, z_{f3} \big), \qquad z_{f1},\, z_q,\,  \tilde{\lambda}_{u},\, \tilde{\lambda}_{d} \in \mathbb{R}_0^+, \\
	z_\ell,\, z_{f2},\, z_{f3},\, \lambda_{ u},\, \lambda_{d},\, \kappa_{f}^p \in\mathbb{C}.
	\end{gathered}
	\end{equation}
This leaves an accidental $ \U(1)_B \times \U(1)_L $ global symmetry, which is preserved by the remaining marginal parameters of the scalar potential.\footnote{On the surface, it would seem that there exists an operator $ R_d R_d R_d H^\ast $~\cite{Arnold:2012sd}; however, this operator vanishes trivially when $ R_d $ has no flavor indices~\cite{Murgui:2021bdy, Dorsner:2022twk, Crivellin:2021ejk}. There is no fully antisymmetric contraction combination of three doublet indices. } The baryon number symmetry ensures the absence of proton decay even with the introduction of LQs, which are infamous for giving rise to $ B $ violation. This was one of the hallmarks of LQ models with gauged lepton number symmetries~\cite{Davighi:2020qqa, Greljo:2021xmg, Davighi:2022qgb}.

\subsection{Producing SM flavor parameters} 
\label{sec:numerical}
We now turn to low energies and see how the SM fermion masses and the CKM matrix are generated. The model reproduces the effective Yukawa couplings of the SM(EFT):
	\begin{equation} \label{eq:SM_Yukawas}
	\L_{\sscript{SMEFT}} \supset\, -Y_u^{pr} \, \overline{q}^p \widetilde{H} u^r -Y_d^{pr} \, \overline{q}^p H d^r -Y_e^{pr} \, \overline{\ell}^p H e^r + \text{H.c.}\,.
	\end{equation}
It generates hierarchical structures for the Higgs Yukawa couplings, $ Y_f $, that allow for a (mostly) perturbative diagonalization. The result is, not only, hierarchical masses but also hierarchical rotation matrices, including the CKM matrix. This is the essence of our solution to the flavor puzzle.  

In addition to the tree-level and loop contributions to the effective Higgs Yukawa couplings in Figs.~\ref{fig:tree-level_masses} and~\ref{fig:loop-level_masses}, there are many other loop contributions formally of the same order. They fall into two categories: $i)$ leg corrections to one of the external states or $ii)$ vertex corrections to the Yukawa couplings. In either case, such contributions shift the UV couplings, but they do not contribute new structures that can lift the rank of the Higgs Yukawa couplings. We will omit all such contributions here.

The quark--Higgs coupling matrices are of the form
    \begin{equation} \label{eq:quark-Higgs_yukawas}
    Y_{u(d)} = \begin{pmatrix}
        b_q \tilde{\lambda}_{u(d)} z_{u(d)} \\
        a_q y_{u(d)} + b_q \lambda_{u(d)} z_{u(d)} \\
        x_{u(d)}
        \end{pmatrix},
    \end{equation}
where 
    \begin{equation}
    a_q = \dfrac{y_q v_\Phi}{M_Q}, \qquad b_q = \dfrac{z_q}{16\pi^2} \dfrac{v_\Phi}{M_L} \! \left(\log \dfrac{M_L^2}{\mu^2} -1 \right),
    \end{equation}
are the suppression factors associated with tree-level and loop-generated operators, respectively. 
The expression for the loop factor $ b_q $ (calculated with the \texttt{Matchete} package~\cite{Fuentes-Martin:2022jrf}) is based on the assumption that $ M_{L,Q} \gg M_{S,R_u,R_d} $, and the renormalization scale $ \mu $ is roughly the mass scale of the LQs. The logarithm thus captures the leading RG running between the masses of the VLFs and the LQs. The mechanism is expected to work also for $ M_{L,Q} \sim M_{S,R_u,R_d} $, but the loop function would be more complicated in this event. 

With singular value decomposition, the Yukawa coupling matrices factor as 
    \begin{equation} \label{eq:Yukawa_SVD}
    Y_f = L_f \widehat{Y}_f R_f^\dagger,
    \end{equation}
where $ \widehat{Y}_f $ is diagonal and $ L_f, R_f $ are unitary rotation matrices. In the quark sector, the decomposition can be performed entirely perturbatively, the full details of which are reported in Appendix~\ref{app:perturbative_diagonalization}. 
The left-handed rotation matrices are hierarchical and are found to satisfy 
    \begin{equation} \label{eq:perturbative_L_approx}
    [L_f]_{ij (ji)} = \mathcal{O}(1) \dfrac{\widehat{Y}^{ii}_{f}}{\widehat{Y}^{jj}_{f}}, \qquad i \leq j, \quad f\in \{ u, d\},
    \end{equation}
where the order-1 coefficient is typically the ratio of two Yukawa couplings. Save for the 1--2 components, the right-handed rotations are even smaller.  

The CKM matrix resulting from the model is nothing but the product of the left-handed quark rotation matrices. Thus, it inherits the hierarchical structure of the effective Higgs Yukawa couplings~\eqref{eq:quark-Higgs_yukawas}. The leading order contribution to each entry in the CKM is given by 
\begin{widetext}
	\begin{equation} \label{eq:ckm_approx}
	\begin{split}
	V_\sscript{CKM} = L_u^\dagger L_d 
	\simeq \begin{pmatrix}
		1 & \big[ \tfrac{m_d z_{d2}}{ m_s z_{d1}} - \tfrac{m_u z_{u2}}{m_c z_{u1}} \big] &  
		\big[ \tfrac{m_d z_{d3}}{m_b z_{d1}} - \tfrac{m_s m_u }{m_b m_c} \tfrac{y_{d3} z_{u2}}{y_{d2} z_{u1}} \big] \\
		\big[\tfrac{m_u z^\ast_{u2}}{m_c z_{u1}} - \tfrac{m_{d} z^\ast_{d2}}{m_s z_{d1}}\big] & 1 & 
		\big[\tfrac{m_s y_{d3}}{m_b y_{d2}} - \tfrac{m_c y_{u3}}{m_t y_{u2}} \big] \\
		\big[\tfrac{m_d}{m_b} \tfrac{y_{d3} z^\ast_{d2}}{y_{d2} z_{d1}} - \tfrac{ m_d z^\ast_{d3}}{m_b z_{d1}} - \tfrac{m_d m_c}{m_s m_t} \tfrac{ z^\ast_{d2} y^\ast_{u3}}{z_{d1} y_{u2}} \big] &
		\big[\tfrac{m_c y^\ast_{u3}}{m_t y_{u2}} - \tfrac{m_s y_{d3}}{m_b y_{d2}} \big] & 1
	\end{pmatrix}\,.
	\end{split}
	\end{equation}
\end{widetext}
This structure easily reproduces the measured values. 

The effective lepton--Higgs coupling matrix produced by the model is
    \begin{equation}\label{eq:ElectronYuk}
    Y_e = \begin{pmatrix}
        \tilde{a}_\ell y_e + b_\ell \lambda^\ast_{ u} z_e \\
        a_\ell y_e - b_\ell \tilde{\lambda}^\ast_{u} z_e \\
        x_e
        \end{pmatrix},
    \end{equation}
where
    \begin{equation}
     \parwidetilde{a}_\ell = \dfrac{\parwidetilde{y}_\ell v_\Phi}{M_L}, \qquad 
    b_\ell = \dfrac{3 z_\ell}{16\pi^2} \dfrac{v_\Phi}{M_Q} \! \left(\log \dfrac{M_Q^2}{\mu^2} -1 \right).
    \end{equation}
The structure is largely identical to the quark sector, with the main difference being that flavor rotations cannot make $ \tilde{y}_q$ and $ \tilde{y}_\ell $ vanish simultaneously. As a consequence, the diagonalization of the matrix involves an order-1 rotation between the light generations. We parametrize $ L_e = V_\ell L_e'$, where $ L_e' $ is the perturbative part of the diagonalization, which follows Eq.~\eqref{eq:perturbative_L_approx}, while $ V_\ell $ capture the large rotation. 

A numerical benchmark point fitting the observed quark and charged lepton masses and the CKM mixing is identified in Appendix~\ref{app:perturbative_diagonalization}. In particular, $ M_{Q,L}/ v_\Phi  = 100 $ is a good starting point for producing the hierarchy of the SM masses. Indeed, one easily finds a benchmark where most of the involved marginal parameters in the UV theory are $\mathcal{O}(0.3)$ except for two (accidentally) smaller parameters contributing to $y_\tau$ and $y_b$.\footnote{One could envisage a 2HDM version of the model explaining this feature. The challenge here is to find a suitable symmetry structure to avoid flavor bounds.} 
We expect that the CKM matrix~\eqref{eq:ckm_approx} is dominated by the down-type contributions, as the hierarchy in the down quark sector is compressed compared to the up quark sector, meaning that the ratio between successive masses is larger. This is reflected in our numerical benchmark.  

Not only do the Higgs couple to all three generations of left-handed fermions after integrating out the VLFs; but the LQ couplings to the SM fields also receive matching contributions. For instance, integrating out $ Q $ at the tree level gives rise to the operators $ \parwidetilde{\Phi}^\alpha \overline{q}_\alpha e R_u $. After $ \Phi $ gets a VEV, all the tree level--generated dimension-5 couplings result in couplings to second-generation fermions suppressed by $ v_\Phi / M_{Q,L} $. Such contributions are similar in size to what is generated from the Yukawa couplings~\eqref{eq:lag_LQ_SM_fermions} by applying left-handed rotations to go to the mass eigenbases. Loop-level contributions will further populate the effective LQ couplings to the SM fermions, but these contributions are suppressed at the level of the 1--3 mixing.

All told, after integrating out the VLFs and taking $ \Phi$ to its VEV, the resulting EFT for the LQs contain the Yukawa couplings
    \begin{equation}
    \L_\sscript{EFT} \supset - \kappa_{u}^{\prime pr}\, \overline{\ell}^p \widetilde{R}_u u^r - \kappa_{d}^{\prime pr}\, \overline{\ell}^p \widetilde{R}_{d} d^r - \kappa_{e}^{\prime pr}\, \overline{q}^p R_u e^r + \text{H.c.}~.
    \end{equation}
The phenomenology of the LQs is primarily determined by the couplings to the mass states of the fermions given by, e.g., 
    \begin{equation} \label{eq:LQ_couplings_mass_states}
    \hat{\kappa}_{eu}^{pr} \equiv (L_e^\dagger \kappa_{u}^{\prime} R_u)^{pr}, \qquad \hat{\kappa}_{de}^{pr} \equiv (L_d^\dagger \kappa_{e}^{\prime} R_e)^{pr}, \qquad \text{etc.}
    \end{equation}
Due to the suppressed couplings to the light generations, we naively expect $ \hat{\kappa}_{eu}^{pr} \sim (L_e^{\prime 3 p})^\ast \kappa_u^r $, $ \hat{\kappa}_{de}^{pr} \sim (L_d^{ 3 p})^\ast \kappa_e^r $, and so forth.  

\section{Phenomenology} 
\label{sec:pheno}

Before delving into the phenomenology of the model, it is crucial to highlight that it exhibits a decoupling limit. By taking the new mass thresholds substantially heavy---given a fixed $ v_\Phi  / M_{Q,L}$ and ensuring $M_{S,R_d,R_u} \lesssim M_{Q,L}$---the SM flavor structure remains unaltered. This might extend almost up to the Planck scale. The decoupling limit detaches the model from experiments, rendering it less pertinent. Conversely, the low-scale variant of the model presents a plethora of interesting phenomenological implications, potentially detectable in current and upcoming experiments. Additionally, the inescapable fine-tuning of the Higgs potential, often called the little hierarchy problem, requires the scales to be as close to the electroweak scale as possible. Subsequent sections will focus on $i)$ the bounds on the new mass thresholds given the current data, and $ii)$ which observables, signatures, and deviation patterns should be prioritized for scrutiny.

As already pointed out, baryon number is an accidental symmetry of the model, which allows this discussion in the first place. The violation of other (approximate) accidental symmetries of the SM is expected to provide the leading constraints. Those include rare flavor-changing neutral current (FCNC) processes in the quark sector, charged lepton flavor violation (cLFV), and CP violation through electric dipole moments (EDM). In the following, we survey these effects, organizing our discussion by the involved fields.

\subsection{Flavored gauge bosons} 
\label{sec:ZpPheno}
The gauge bosons of \gf become massive as $ \Phi $ develops a VEV $v_\Phi$ at high scales. In the simplest case, with no other symmetry-breaking scalars, this results in a heavy, degenerate vector triplet $ Z^a_\mu $, which carries the flavor of the light generations but is otherwise neutral. The phenomenology of flavored gauge bosons from $ \SU(2) $ flavor groups was studied in detail in Ref.~\cite{Darme:2023nsy}, and here we only consider the most important bounds for our model.

In spirit with our approach to marginal couplings, we consider only the case of a reasonably large gauge couplings $ g_f \gtrsim \SI{10}{TeV}/M_Z $, in which case the gauge bosons are too heavy to be produced on-shell in current collider experiments. The low-energy physics impact of these flavored gauge bosons is due to four-fermion operators generated at the tree level. Using Fierz identities for the \gf generators, we find (Greek indices take values in the light generations only, $ \alpha, \beta, \ldots \in \{1,\, 2\} $)
    \begin{multline} \label{eq:4-fermion_gauge_basis}
    \L_{\sscript{SMEFT}} \supset - \dfrac{1}{2 v_\Phi^2} \big[\delta\ud{\alpha}{\delta} \delta\ud{\gamma}{\beta} - \dfrac{1}{2} \delta\ud{\alpha}{\beta} \delta\ud{\gamma}{\delta} \big] \big[(\overline{q}_\alpha \gamma^\mu q^\beta) (\overline{q}_\gamma \gamma^\mu q^\delta) \\  + 2 (\overline{q}_\alpha \gamma^\mu q^\beta) (\overline{\ell}_\gamma \gamma^\mu \ell^\delta) + (\overline{\ell}_\alpha \gamma^\mu \ell^\beta) (\overline{\ell}_\gamma \gamma^\mu \ell^\delta)\big].
    \end{multline}
As per usual, the low-energy physics of the massive vector is independent of the gauge coupling and solely depends on the VEV $v_\Phi $. 

For completeness, let us comment that the flavon field (the radial component of $\Phi$) is expected to have a mass of order $v_\Phi$ and suppressed couplings to SM fermions. Therefore, this field does not produce any observable phenomenology in present experiments.

\paragraph{Meson mixing} One might anticipate strong bounds from $ \Delta F =2 $ processes contributing to meson--anti-meson mixing; however, it turns out that there is a strong GIM-like mechanism protecting flavor transitions within the same spices.
As an example, we consider the contribution to kaon mixing, which in the mass basis of the quarks, read   
    \begin{equation}
    \L_{\sscript{LEFT}} \supset - \dfrac{1}{4v_\Phi^2} A_{sd}^2 (\overline{s}_\LL \gamma_\mu d_\LL)^2, 
    \end{equation}
where the coupling matrix
    \begin{equation}
    A_{f_p f'_r} = \big[ L_{f}^\dagger \diag(1,\, 1,\, 0) L_{f'} \big]_{pr}.
    \end{equation}
is determined by the left-handed rotation matrices. The gauge basis interaction is proportional to the identity but for the third generation, which is singlet. Unitarity of the rotation matrices implies that flavor violation within the same species must go through the third generation:  
    \begin{equation} \label{eq:suppresion_mechanism}
    A_{f_p f_r} = \delta_{pr} - [L_f^\ast]_{3p} [L_f]_{3r}, \qquad f=f'.
    \end{equation}
 
In the example of kaon-mixing, the flavor matrix is 
    \begin{equation}
    A_{sd} =  \dfrac{a_q b_q \tilde{\lambda}_{d} y_{d3}}{x_{d3}^2} \left( \dfrac{y_{d3} z_{d2}^\ast}{y_{d2}} - z^\ast_{d3}\right) \sim \dfrac{m_d m_s}{m_b^2} \sim \num{e-5}.
    \end{equation}
Accordingly, there is no significant contribution to kaon mixing. The same is true for $ D -\overline{D}$ mixing, where $ A_{cu} $ is even smaller due to the larger hierarchy in the up quark sector. 

If we assume that \eqref{eq:perturbative_L_approx} provides a good approximation of the left-handed mixing matrices, the strongest meson mixing bound will come from $ B_s -\overline{B}_s $ mixing. A rough comparison with current bounds (neglecting RG effects) yields~\cite{Silvestrini:2018dos} 
    \begin{equation}
    \dfrac{|\Re A^2_{sb}|}{0.02^2} \left( \dfrac{\SI{10}{TeV}}{v_\Phi} \right)^{\!\! 2} \lesssim \num{22}.
    \end{equation}
Even with, as we shall see, a very small $ v_\Phi \sim \SI{10}{TeV} $, the naive expectation of $ A_{sb} $ would not conflict with the bound. The GIM-like suppression mechanism applies to $ \mu \to 3 e $ decays too, which are suppressed by $ A_{\mu e} $ and similarly irrelevant. 


\paragraph{Lepton flavor--violating kaon decays}
A stringent constraint on the symmetry-breaking scale comes from bounds on $ K_L \to \mu e $. This is an example of a flavor-transferring process~\cite{Darme:2023nsy}, connecting light flavors of one species (down-type quarks) to those of another (charged leptons). Such processes mediated by the flavored gauged bosons are entirely unsuppressed.

The heavy vectors contribute to the LEFT operators~\cite{Jenkins:2017jig} (cf. Eq.~\eqref{eq:4-fermion_gauge_basis})\footnote{We do not include RG effects in this leading order analysis although they could be sizable.}
	\begin{equation}
	\begin{split}
	\L_{\sscript{LEFT}} \supset &- \dfrac{1}{v_\Phi^2} A_{se} A_{\mu d} (\overline{\mu}_\LL \gamma_\mu e_\LL)(\overline{s}_\LL \gamma_\mu d_\LL)  \\
	&- \dfrac{1}{v_\Phi^2}  A_{s\mu} A_{e d} (\overline{e}_\LL \gamma_\mu \mu_\LL)(\overline{s}_\LL \gamma_\mu d_\LL) + \mathrm{H.c.}\,,
	\end{split}
	\end{equation}
which give rise to $ K_L \to \mu e $ decays. We have omitted contributions proportional to $ A_{e\mu} A_{sd} = A^\ast_{\mu e} A^\ast_{d s} $, which are severely suppressed. 
Based on these LEFT contributions, we find a total branching ratio of~\cite{Marzocca:2021miv,Angelescu:2020uug} 
    \begin{multline} \label{eq:BR_KL->mue}
    \mathrm{BR}\big(K_L \to \mu^{\pm} e^{\mp} \big) \\= \num{5.9e-12} \left( \dfrac{\SI{300}{TeV}}{v_\Phi} \right)^{\!\! 4}  
    \big| A_{se} A_{\mu d} + A_{de} A_{\mu s} \big|^2\,.
    \end{multline}
The experimental bound set by BNL is $ \mathrm{BR}\big(K_L \to \mu^{\pm} e^{\mp} \big) < \num{4.7e-12} \;@\; 90\% \text{CL} $~\cite{BNL:1998apv}.

The combination of coupling matrices appearing in Eq.~\eqref{eq:BR_KL->mue} is expected to be $ \lesssim 1$, and we would generally expect a bound $ v_\Phi \gtrsim \SI{300}{TeV} $. However, the bound can be avoided entirely for some rotation matrices, due to the $ \mathcal{O}(1) $ rotation $ V_\ell $ of the left-handed leptons. To illustrate this point, we consider a simple scenario, where $ L_u = L_e' = \mathds{1} $. In this scenario, the CKM matrix is produced by the down-type rotation and 
    \begin{equation} \label{eq:mixing_scenario}
    A_{d_p \ell_r} = A^\ast_{\ell_r d_p} = \big[V_\sscript{CKM} \diag(1,1,0) V_\ell \big]_{pr}\,.
    \end{equation}
With these assumptions
    \begin{equation} \label{eq:kaon_LFV_factor}
    A_{de} A_{\mu s} + A_{d\mu} A_{se} = 0.90 \,c_{2\ell} + 0.44\, s_{2\ell}\,,
    \end{equation}
where the trigonometric functions $ c_{2\ell}$ and $ s_{2\ell} $ are defined in Eq.~\eqref{eq:def_Vl}. The bound on $ v_\Phi $ from the non-observation of lepton flavor--violating kaon decays is entirely avoided for $ s_\ell = -0.53 $; however, it turns out that a complementary bound from muon conversion is sufficient to eliminate the possibility of a low scale $ v_\Phi $.

\paragraph{Muon conversion on heavy nuclei}
Another strong bound comes from muon conversion on nuclei. This process also involves a flavor-transferring processe, so there is no suppression from the GIM-like mechanism. The LEFT operators relevant to the process are (cf. Eq.~\eqref{eq:4-fermion_gauge_basis})
    \begin{equation}
    \begin{split}
    \L_\sscript{LEFT} \supset &- \dfrac{1}{v_\Phi^2} A_{ue} A_{\mu u} (\overline{\mu}_\LL \gamma_\mu e_\LL)(\overline{u}_\LL \gamma_\mu u_\LL)  \\
    &- \dfrac{1}{v_\Phi^2}  A_{de} A_{\mu d} (\overline{\mu}_\LL \gamma_\mu e_\LL)(\overline{d}_\LL \gamma_\mu d_\LL) + \mathrm{H.c.}\,.
    \end{split}
    \end{equation} 
To a good approximation $ A_{e\mu} \simeq 0 $, and we have omitted such contributions to the Wilson coefficients.

The conversion ratio of muons to electrons on gold atoms is given by~\cite{Darme:2023nsy}
    \begin{multline} \label{eq:muon_conversion_rate}
    \mathrm{CR}(\mu \mathrm{Au} \to e \mathrm{Au}) = \num{2e-11} \cdot \left( \dfrac{\SI{300}{TeV}}{v_\Phi} \right)^{\!\!4} \\
    \times \left|  A_{e u} A_{u \mu} + 1.143\, A_{e d} A_{d \mu} \right|^2.
    \end{multline}
The present experimental limit of $ \mathrm{CR}(\mu \mathrm{Au} \to e \mathrm{Au})< \num{7e-13} $ is set by SINDRUM-II~\cite{SINDRUMII:2006dvw}. Muon conversion provides a slightly stronger bound than $ \mathrm{BR}\big(K_L \to \mu^{\pm} e^{\mp} \big) $ but also here it is possible to circumvent it for suitable coupling matrices. 

In the scenario of~\eqref{eq:mixing_scenario}, the muon conversion rate~\eqref{eq:muon_conversion_rate} becomes 
    \begin{multline}
    \mathrm{CR}(\mu \mathrm{Au} \to e \mathrm{Au}) = \num{2e-11} \cdot \left( \dfrac{\SI{300}{TeV}}{v_\Phi} \right)^{\!\!4} \\
    \times \left|  1.01 \,s_{2\ell} -0.25 \,c_{2\ell} \right|^2.
    \end{multline}
This is out of phase with combination~\eqref{eq:kaon_LFV_factor}, so the bounds cannot be simultaneously avoided. Between them they constrain $ v_\Phi > \SI{300}{TeV} $ in the low end ($ \theta_\ell = 1.66 $) and $ v_\Phi > \SI{700}{TeV} $ at the high end ($ \theta_\ell = 0.97 $). Our benchmark of $ v_\Phi = \SI{1}{PeV} $ passes the bound irrespective of the angle. 

The planned MU2e and COMET experiments projects a future sensitivity of $ \mathrm{CR}(\mu \mathrm{Au} \to e \mathrm{Au})$ at the level of $ \num{e-17}$~\cite{Bernstein:2019fyh,Moritsu:2022lem}. The use of aluminum targets rather than the gold target employed by SINDRUM-II means changing nuclear form factors and muon capture rates in our analysis, and the comparison with the conversion rate on gold is not one-to-one. Nevertheless, we expect an improved sensitivity of almost an order of magnitude on $ v_\Phi $ and an exciting possibility of discovering indirect signs of the flavored vector bosons.

\subsection{Vectorlike fermions} 
\label{sec:VLFpheno}
We have seen that a realistic SM flavor structure is realized when $ M_{Q, L} \sim 100 v_\Phi $. Even for an absurdly low symmetry-breaking scale, the hierarchy implies VLF masses at the PeV scale. Consequently, only the most sensitive low-energy observables can hope to place meaningful bounds on the VLF masses.  

The LEFT dipole operator for the charged leptons can be parametrized by
    \begin{equation} \label{eq:LEFT_dipole}
    \L_\sscript{LEFT} = - e\, v_\sscript{EW}\, C_{e\gamma}^{pr}\, F^{\mu\nu} \overline{\ell}^p  \sigma_{\mu\nu} P_\RR \ell^r\,,  
    \end{equation}
where $ v_\sscript{EW} = \SI{174}{GeV} $ is the Higgs VEV.
Contributions to this operator are generated by diagrams similar to the loop contribution to the Higgs--Yukawa coupling from the charged leptons, Fig.~\ref{fig:loop-level_masses}, but dressed with gauge bosons. The responsible loop is evaluated using the \texttt{Matchete} package~\cite{Fuentes-Martin:2022jrf}, and we find that in the mass basis
    \begin{equation}
    C_{e\gamma}^{11} = \dfrac{\widehat{Y}_{e}^{11}}{3 M_Q^2} \dfrac{\log(M_Q/\mu) - \tfrac{11}{8}}{\log(M_Q/\mu) - \tfrac{1}{2}} \! \left(\!1 - \dfrac{b_{\ell} \lambda_e z_{e2}}{A_{\ell} y_{e2}} +\ldots \!\right),
    \end{equation}
where $ \mu \ll M_Q $ is of the order of the LQ masses.\footnote{This formula does not include the EM running of the dipole operator between $ \mu $ and $ m_e $.}  

In terms of the perturbative diagonalization of the Higgs--lepton Yukawa, the leading contribution to the electron dipole coefficient is aligned with the electron coupling $ \widehat{Y}^{11}_e $. The subleading term contributes with a phase relative to electron mass but is suppressed by roughly $ m_e / m_\mu $ compared to the leading term. The resultant contribution to the lepton EDM is~\cite{Jegerlehner:2009ry}
    \begin{equation} \label{eq:VLF_EDM_contribution}
    d_e \simeq \num{3.2e-32} e \,\si{cm} \cdot \left(\dfrac{\SI{1}{PeV}}{M_Q} \right)^{\!\! 2} \Im \!\left[ \dfrac{b_\ell}{|b_\ell|} \dfrac{\lambda_e z_{e2}}{\tilde{\lambda}_e z_{e1}} \right]. 
    \end{equation}
This is easily compatible with the experimental limit of $ |d_e^\mathrm{exp}| < \num{1.1e-29} e \,\si{cm} \;@\; 90\% \text{CL}$ reported by the ACME experiment~\cite{ACME:2018yjb}, even for the smallest conceivable $ M_Q$ masses. The limits set by $ \mathrm{Br}[\mu \to e \gamma] $ are even weaker, and we will not report them here.

\subsection{Leptoquarks} 
\label{sec:LQ_pheno}

The gauged \gf symmetry along with the assumption that marginal couplings cluster around $\mathcal{O}(0.3)$ determines the flavor structure of the LQ interactions (see Section~\ref{sec:numerical} for details). Nonetheless, different mass spectra can alter the resulting phenomenology. Interesting effects can arise when the LQs are light, as allowed for by the mechanism $M_{S, R_{u}, R_{d}} \lesssim M_{Q,L}$. A limitation on how light the LQs can be comes from the scalar potential and its VEV structure. The quartic couplings~\eqref{eq:crossquartics} will destabilize the desired VEV structure unless at least one of the following conditions is met:\footnote{Consider a toy potential $V(x,y,z) = - v \, x y z + x^4 +y^4 +z^4$. In such a configuration all fields would develop a VEV: $\langle x \rangle = \langle y \rangle = \langle z \rangle = v / 4$ (up to a tetrahedral symmetry). When one of the fields gets a mass $ m \geq v /2$, the minimum of the potential moves to the origin, and no symmetry breaking will occur.}
\begin{itemize}
    \item Scenario I: $M_S \gtrsim  v_\Phi $ with $R_d$ and $R_u$ potentially lighter, 
    \item Scenario II: $M_{R_d}, M_{R_u} \gtrsim v_\Phi$ with $S$ potentially lighter.
\end{itemize}
The former scenario predicts deviations in rare flavor transitions, whereas the latter may give rise to distinct collider signatures. In the following, we discuss the two scenarios separately.

The first scenario characterizes a potentially large violation of approximate flavor symmetries if $R_d$ and (or) $R_u$ are sufficiently light. We are interested in the parameter range $v_\sscript{EW} \lesssim M_{R_{d},R_{u}}\lesssim v_\Phi $. The two leptoquarks are $\SU(2)_\LL$ doublets with electromagnetic components $\left(\frac{2}{3},-\frac{1}{3}\right)$ for $R_d$ and $\left(\frac{5}{3},\frac{2}{3}\right)$ for $R_u$, respectively. In principle, LQ states with electromagnetic charge $\frac{2}{3}$, including $ S $, can mix after the $\SU(2)_\LL \times \SU(2)_{q+\ell}$ breaking. However, the constraint on $ v_\Phi $ from Sec.~\ref{sec:ZpPheno} and the consistency condition $M_S \gtrsim v_\Phi$ together imply that the mixing due to Eq.~\eqref{eq:crossquartics} is suppressed. Another potential source of mixing, $H H R_d R^\ast_u$ is negligible compared to the LQ mass terms given the limits derived below. Therefore, unless accidentally $ \big| M_{R_d}^2 -  M^2_{R_u} \big|\lesssim v_\sscript{EW}^2$, the expected mixing is small and its effects omitted in the following. Thus, we can discuss the phenomenology of each LQ in isolation.

\subsubsection{Scenario I: The phenomenology of a light $R_d$} 
\label{sec:RdPheno}

Integrating out the $R_d$ leptoquark at the tree level gives
\begin{equation}\label{eq:lag_EFT_Rd}
    \mathcal{L}_{\sscript{ SMEFT}} \supset - \frac{\kappa^{p*}_d \kappa^{r}_d}{2 M^2_{R_d}} \big( \overline{\ell}^3 \gamma_\mu \ell^3 \big) \big( \overline{d}^p \gamma^\mu d^r \big)~,
\end{equation}
where $p,r=1,2,3$ and $\kappa^{p}_d$ are Yukawa couplings to SM fermions~\eqref{eq:lag_LQ_SM_fermions}. 
As discussed at the end of Section~\ref{sec:numerical}, there are additional contributions when integrating out VLF. We expected such contributions to be somewhat smaller than those generated by the perturbative rotations to the mass basis, which are enhanced due to the accidentally small $ \tau $ Yukawa.
To illustrate the general expectations, we keep only the effects of rotations where appropriate in the following. 

\begin{figure}[t]
        \centering
        \includegraphics[width=0.5\textwidth]{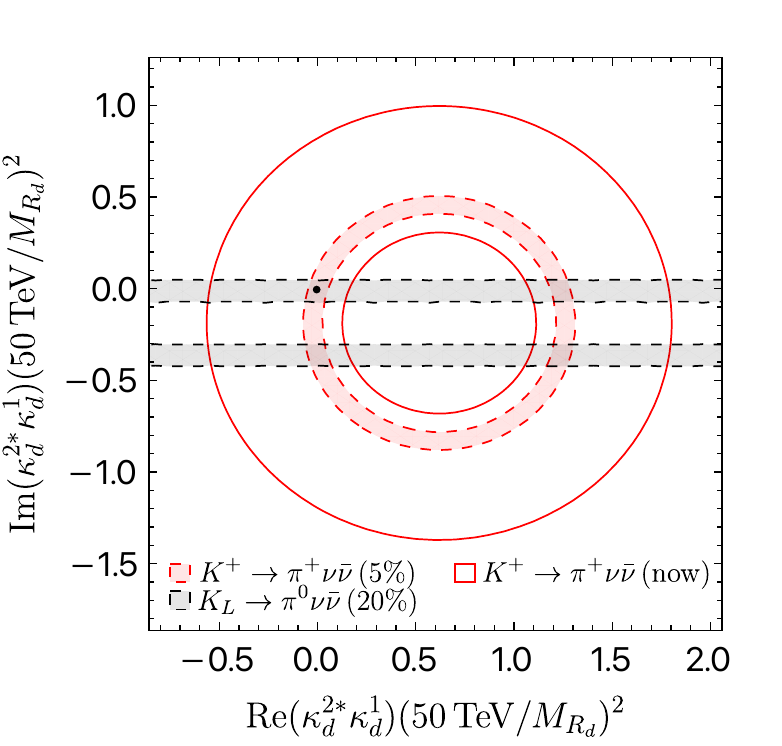}
        \caption{Constraints on $R_d$ leptoquark from rare kaon decays. The most recent NA62 analysis~\cite{NA62:2021zjw} confines the region satisfying Eq.~\eqref{eq:NA62} within the red solid boundaries. KOTO~\cite{KOTO:2018dsc} does not restrict any part of this range. Shaded regions within dashed lines represent anticipated boundaries from future facilities, see~\cite{NA62KLEVER:2022nea, HIKE:2022qra}.}
        \label{fig:kaon}
\end{figure}

\paragraph{Rare Kaon decays to neutrinos} The effective Lagrangian~\eqref{eq:lag_EFT_Rd} modifies rare FCNC decays $d^p \to d^r \nu_\tau \bar \nu_\tau$. In particular, measurements of rare kaon decays:
\begin{align}
      \mathrm{BR}(K^+ \to \pi^+ \nu \bar \nu ) &=  10.6^{+4.1}_{-3.5}\cdot 10^{-11}\quad \text{(NA62~\cite{NA62:2021zjw})}\,,\label{eq:NA62}\\
  \mathrm{BR}(K_L \to \pi^0 \nu \bar \nu ) & < \num{3e-9}\,\text{at 90\%\,CL}\quad \text{(KOTO~\cite{KOTO:2018dsc})}\,,
\end{align}
set the most stringent limits. Using expressions from Refs.~\cite{Buras:2020xsm, Brod:2021hsj}, we set limits on the corresponding coupling and mass combinations as shown in Fig.~\ref{fig:kaon}. The present limits are shown within red solid lines. The future kaon physics program~\cite{NA62KLEVER:2022nea, HIKE:2022qra} has the potential to improve by almost one order of magnitude. Dashed lines show the improvement assuming $5\%$ and $20\%$ precision for the charged and neutral modes, respectively. In particular, for $\kappa^1_d = \kappa^2_d \simeq 0.3$, the projected sensitivity will reach masses of about $50$\,TeV.

\paragraph{Rare $B$ decays to neutrinos} Another important channel for $ R_d $ is $B \to K^{(*)} \nu \bar \nu$ decays~\cite{Belle:2017oht, Belle-II:2021rof, GlazovEPS} which probe a different combination of leptoquark couplings, namely $\kappa^{2 \ast}_d \kappa^3_d$. The branching ratio for the $K^*$ mode normalized to the SM prediction is bounded to 
\begin{equation}\label{eq:RkSnu}
    R^\nu_{K^*} < 2.7\,\text{at 90\%\,CL}\quad\text{(Belle~\cite{Belle:2017oht})}\,.
\end{equation}
Instead, a recent major update from the Belle~II collaboration on the $K$ mode reports evidence for the observation~\cite{GlazovEPS}. Moreover, the reported rate is about $2.8\sigma$ above the SM prediction (for interpretations see~\cite{Allwicher:2023syp, Bause:2023mfe, Felkl:2023ayn}). When combined with previous $B \to K \nu \bar \nu$ measurements, one finds~\cite{GlazovEPS}
\begin{equation}\label{eq:RKnu}
    R^\nu_{K} = ~ 2.8 \pm 0.8 \,,
\end{equation}
for the branching ratio normalized to the SM prediction. 

\begin{figure}[t]
         \centering
        \includegraphics[width=0.48\textwidth]{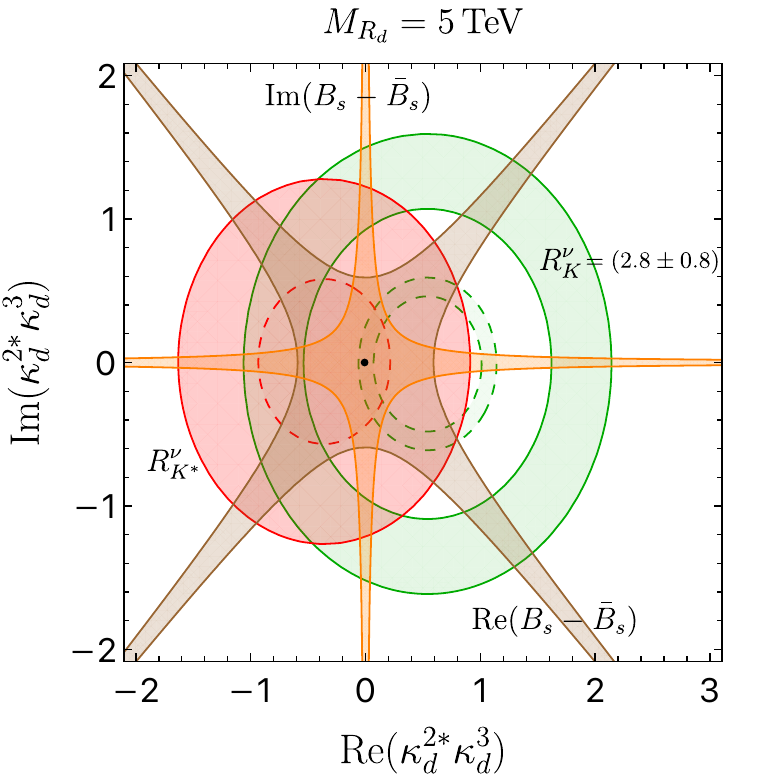}
        \caption{Constraints on $R_d$ leptoquark from $B \to K^{(*)} \nu \bar \nu$ decays and $B_s - \bar B_s$ oscillations. The leptoquark mass is set to $M_{R_d} = \SI{5}{TeV}$. Shown with solid green is the latest $R^\nu_K$ average from Eq.~\eqref{eq:RKnu}, including the most recent Belle II measurement~\cite{GlazovEPS}. Solid red is for $R^\nu_{K^*}$ and satisfies Eq.~\eqref{eq:RkSnu}. Brown and orange show the constraints from $B_s - \bar B_s$ oscillations, namely Eq.~\eqref{eq:Bsmix}.  The dashed lines show the future Belle II projections with $50$\,ab$^{-1}$~\cite{Belle-II:2022cgf}. }
        \label{fig:BKnunu}
\end{figure}

Taking expressions for $ R^\nu_{K^{(*)}}$ from~\cite{Buras:2020xsm,Buras:2014fpa}, we show the preferred parameter space in Fig.~\ref{fig:BKnunu}. Fixing $M_{R_d} = 5$\,TeV, the relevant magnitude of the couplings is $|\kappa^{2*}_d \kappa^3_d| = \mathcal{O}(1)$. Hence, $B$ decays overcome $K$ decays and provide the leading phenomenology only for a somewhat small coupling ratio $|\kappa^{1}_d / \kappa^{3}_d | \lesssim 10^{-2}$. 

Future Belle II data has a great potential for further improvements. Assuming the SM rates and the improved projections for $5$\,ab$^{-1}$, the $K\,(K^*)$ mode will be measured to $19\% \,(40\%)$ precision~\cite{Belle-II:2022cgf} (see also~\cite{Belle-II:2018jsg}). This will further improve by a factor $\sim 2$ at $50$\,ab$^{-1}$, namely to $8\% \,(23\%)$. These ultimate projections are shown with dashed lines in Fig.~\ref{fig:BKnunu}.
Slightly weaker, but otherwise similar, constraints are derived from $B \to \pi \nu \bar \nu$ and $B \to \rho \nu \bar \nu$ on $\kappa^{1*}_d \kappa^3_d$ combination~\cite{Belle:2017oht}.

\paragraph{Meson mixing} Matching the $R_d$ leptoquark at one-loop order to the SMEFT contributes~\cite{Dorsner:2016wpm, Crivellin:2021lix}
\begin{equation}
    \mathcal{L}_{{\rm SMEFT}} \supset - \frac{\kappa^{p\ast }_d \kappa^{r}_d \kappa^{s \ast }_d \kappa^{t}_d }{64\pi^2 M^2_{R_d}} \big(\overline{d}^p \gamma^\mu d^r \big) \big( \overline{d}^s \gamma^\mu d^t \big)\,.
\end{equation}
This induces contributions to neutral meson oscillations. Assuming either real or imaginary $(\kappa^{p*}_d \kappa^{r}_d)^2$ independently, we adopt the $95\%$\,CL bounds in the SMEFT from~\cite{Silvestrini:2018dos} extrapolated to $\mu=1$\,TeV (see also~\cite{Aebischer:2020dsw, Isidori:2010kg}). From $K- \overline{K}$ oscillations, we find
\begin{align}
    \left| \Re\! \big[ (\kappa^{2*}_d \kappa^1_d)^2 \big] \right| \left(\frac{50\,\text{TeV}}{M_{R_d}}\right)^{\!\! 2} &\lesssim 1.0 \quad(\Delta m_K)~,\\
    \left| \Im\! \big[ (\kappa^{2*}_d \kappa^1_d)^2 \big] \right| \left(\frac{50\,\text{TeV}}{M_{R_d}}\right)^{\!\!2} &\lesssim \num{3e-3} \quad (\epsilon_K)~.
\end{align}
Clearly, $\epsilon_K$ is the most stringent present limit on the $R_d$ leptoquark for $\mathcal{O}(1)$ couplings. However, meson decays and meson mixings depend differently on the couplings (quadratic versus quartic). For light leptoquarks and small couplings, kaon decays will give a relevant bound. For example, setting $M_{R_d} = 5$\,TeV and $\kappa^{2}_d\simeq 0.3$, one finds that $|\kappa^{1}_d|\simeq \mathcal{O}(0.01)$ passes the $\epsilon_K$ bound, as well as, $K\to \pi \nu \nu$ which becomes competitive. Interestingly, even for larger masses $M_{R_d} = 50$\,TeV and couplings $\kappa^{p}_d \sim 0.3$, future kaon decays become competitive to $\epsilon_K$.

Moving on to the $B$ physics of $B- \overline{B}$ and $B_s- \overline{B}_s$ oscillations, we find~\cite{Silvestrini:2018dos}
\begin{align}
    \left| \Re(\Im)\! \big[(\kappa^{1*}_d \kappa^3_d)^2 \big] \right| \left(\frac{50\,\text{TeV}}{M_{R_d}} \right)^{\!\!2} &\lesssim 1.6 \,(1.4)~,\\
    \left| \Re(\Im) \! \big[(\kappa^{2*}_d \kappa^3_d)^2\big] \right| \left(\frac{5\,\text{TeV}}{M_{R_d}}\right)^{\!\!2} &\lesssim 0.35 \, (0.12).~\label{eq:Bsmix}
\end{align}
As an illustration, we impose the last equation on Fig.~\ref{fig:BKnunu}. Regions colored brown and orange satisfy the inequalities from Eq.~\eqref{eq:Bsmix} for the real and imaginary parts, respectively. It is instructive to compare these bounds against $B\to K^{(*)} \nu \bar \nu$. In the case of $ M_{R_d} = \SI{5}{TeV}$, $\Re(\kappa^{2*}_d \kappa^3_d) = -0.5$ and small imaginary part can barely accommodate for $R^\nu_K$ anomaly reported in~\cite{GlazovEPS}. For lower masses, the parameter space opens up due to the aforementioned scaling, while $R^\nu_K$ is in tension with $B_s$ mixing for larger masses. This scenario predicting, $M_{R_d} \lesssim \SI{5}{TeV}$, is exciting for collider searches. The present ATLAS and CMS direct searches for scalar LQs exclude masses $\lesssim 1.5$\,TeV~\cite{ATLAS:2023uox, CMS:2018iye}. Although the remaining parameter space is squeezed, it will take a future collider to probe it fully~\cite{Azatov:2022itm}.

\paragraph{Muon to electron conversion} The perturbative left-handed rotation in the lepton sector will transmit the interactions to light leptons. Due to accidentally small $y_\tau$, the mixing angles can be rather large. As a benchmark, in the following we consider $ |L_e^{32}| = 0.1 $ and $ |L_e^{31}| = \num{e-3} $ in agreement with Section~\ref{sec:numerical}. This leads to the charged lepton flavor violation. In particular, the leading bound is the $\mu \to e$ conversion on heavy nuclei. Translating the limit from~\cite{Davidson:2020hkf}, we find
\begin{equation}
    | \kappa^{1}_d |^2  \frac{|L_e^{32}|}{0.1} \; \frac{|L_e^{31}|}{\num{e-3}} \left(\frac{5\,\text{TeV}}{M_{R_d}}\right)^{\!\!2} \lesssim 1~.
\end{equation}
As per the discussion in Sec.~\ref{sec:ZpPheno}, the future muon conversion experiments are expected to be able to probe $ M_{R_d} $ masses an order of magnitude beyond the present.

\paragraph{Kaon decays to charged leptons} 
Another important set of transitions includes $s \to d \bar \ell^p \ell^r$ decays. For instance, $\mathrm{BR}(K_L \to \mu \mu)_{{\rm SD}}\lesssim 2.5\times 10^{-9}$~\cite{Isidori:2003ts} implies~\cite{Marzocca:2021miv}
\begin{equation}
        \left| \kappa^{2}_d \kappa^1_d\right| \frac{|L_e^{32}|^2}{0.1^2} \left(\frac{5\,\text{TeV}}{M_{R_d}}\right)^{\!\!2} \lesssim 1~.
\end{equation}
These are not competitive with the neutrino modes shown in Fig.~\ref{fig:kaon}. Potential future improvements by an order of magnitude on the mass will involve another observable, $\mathrm{BR}(K_S \to \mu \mu)_{\ell=0}$~\cite{Dery:2021mct, Brod:2022khx}. 

The lepton flavor-violating channel, $\mathrm{BR}(K_L \to \mu e)\lesssim 4.7\times 10^{-12}$~\cite{BNL:1998apv}, implies even lesser bound due to the suppression from $e$--$\tau$ mixing~\cite{Marzocca:2021miv}
\begin{equation}
        \left| \kappa^{2}_d \kappa^1_d\right| \frac{ |L_e^{32}|}{0.1} \, \frac{ |L_e^{31}|}{\num{e-3}} \left(\frac{5\,\text{TeV}}{M_{R_d}}\right)^{\!\!2} \lesssim 5~.
\end{equation}

\paragraph{$B$-decays to charged leptons} 
Two insertions of the $\mu$--$\tau$ rotation allow for $ b \to s \mu \mu $ decays. The contribution to the lepton flavor universality ratios $R_{K^{(*)}}$~\cite{Hiller:2003js,DAmico:2017mtc} is given by
\begin{align}
    R_K &\simeq 1 + 0.06 \, \Re(\kappa^{2*}_d \kappa^3_d) \frac{|L_e^{32}|^2}{0.1^2} \left(\frac{5\,\text{TeV}}{M_{R_d}}\right)^{\!\!2},\\
    R_{K^*} &\simeq 1 - 0.7 (R_K - 1)\,.
\end{align}
Interestingly, the LHCb measurements in the central $q^2$ bin $R_K \simeq 0.95 \pm 0.05$ and $R_{K^*} \simeq 1.03 \pm 0.07$~\cite{LHCb:2022qnv, LHCb:2022vje} provide a complementary probe to $b \to s \nu_\tau \bar \nu_\tau$. In fact, the aforementioned $R^\nu_K$ anomaly from Eq.~\eqref{eq:RKnu} correlates with $R_K \simeq 0.97$ ($R_{K^*} \simeq 1.02$) for $L_e^{23} =0.1$ and $M_{R_d}=5$\,TeV. The projected sensitivity for $R_K$ and $R_K^*$ after the LHCb upgrade II is $0.7\%$ and $0.8\%$, respectively~\cite{LHCb:2018roe}, which is comparable with the present theory uncertainty~\cite{Bordone:2016gaq, Isidori:2022bzw}.

\subsubsection{Scenario I: The phenomenology of a light $R_u$} 
\label{sec:RuPheno}
In contrast to $ R_d$, all the most important bounds on $ R_u $ comes from loop-level effects. Similarly to the $ R_d $ couplings to left-handed lepton mass eigenstates, we expect that the effective coupling of $ R_u $ to the left-handed leptons to be dominated by the $ L_e $ rotation mixing. We expect it to transfer the marginal $ \kappa_u^p $ coupling from $ \ell^3 $ to the light generation leptons with a minor enhancement due to the small $ \tau $ Yukawa. We disregard the presumed subdominant contributions from higher-dimension operators in the following discussion.

\paragraph{Leptonic dipoles}
The chirally enhanced contribution of $ R_u$ to the leptonic dipoles poses severe bounds on the LQ mass when confronted with the strong complementary experimental limits from electron EDM and $ \mu \to e \gamma $ decay rate. 
In the leading-log approximation the $ R_u $ contribution to the LEFT leptonic dipole~\eqref{eq:LEFT_dipole} at the top scale is\footnote{Again the loop is evaluated with the \texttt{Matchete} package~\cite{Fuentes-Martin:2022jrf}.}
    \begin{equation}
    C_{e\gamma}^{pr} = - \dfrac{1}{16 \pi^2} \dfrac{ (L_e^{3p})^\ast \kappa_{u}^{3} x_{3u}\kappa_e^r }{M_{R_u}^2} \log \dfrac{M_{R_u}^2}{m_t^2}.
    \end{equation}
The internal loop structure is dominated by the LQ couplings to third-generation left-handed quarks.

The off-diagonal components of the leptonic dipole operator give rise to LFV decays, the branching ratio of which is determined with Refs.~\cite{Crivellin:2018qmi,Isidori:2021gqe}. Confronting the current experimental bound $ \mathrm{Br}[\mu \to e \gamma] < \num{4.2e-13}$ at $90\%$\,CL from the MEG collaboration~\cite{MEG:2016leq}, we obtain a bound of 
    \begin{equation} \label{eq:Ru_LFV_constraint}
     |\kappa_{u}^{3} x_{3u} \kappa_e^1 | \, \dfrac{\big| L_e^{32}\big|}{0.1} \! \left( \dfrac{\SI{500}{TeV}}{M_{R_u}} \right)^{\!\! 2} \dfrac{\log \tfrac{M_{R_u}}{m_t}}{8} < \num{0.017} \,.
    \end{equation}
This results in a bound $ M_{R_u} \gtrsim \SI{500}{TeV} $ when the couplings are $ \mathcal{O}(0.3)$. The MEG-II experiment~\cite{MEGII:2018kmf} will improve the branching ratio limit by a factor of seven.

The constraint from the electron EDM is even a little stronger (cf. the discussion around Eq.~\eqref{eq:VLF_EDM_contribution}), clocking in at 
    \begin{equation} \label{eq:Ru_EDM_constraint}
    \dfrac{|x_{3u} \Im( (L_e^{31})^\ast \kappa_u^1 \kappa_e^1)|}{\num{e-3}} \left(\dfrac{\SI{500}{TeV}}{M_{R_u}} \right)^{\!\! 2} \dfrac{\log \tfrac{M_{R_u}}{m_t}}{8}< \num{4e-3} \,.
    \end{equation}
Future improvement of the electron EDM by a factor 10--20 is possible~\cite{EuropeanStrategyforParticlePhysicsPreparatoryGroup:2019qin}. Clearly, it is possible to evade the bound on $ M_{R_u} $ and obtain an unexpectedly light LQ if the combination of couplings in the constraints~\eqref{eq:Ru_LFV_constraint} and~\eqref{eq:Ru_EDM_constraint} are accidentally small. 

One could also consider LFV constraints from $ \mu \to 3 e$ decays; however, these turn out to be much weaker, resulting in  $ M_{R_u} \gtrsim  \SI{10}{TeV} $. These weaker bounds, generated from a box diagram in the UV, depend only on the coupling $ \kappa_e $ and are not strictly correlated with the bounds from the lepton dipoles.

\paragraph{Meson mixing}
The $ R_u $ LQ produces the SMEFT operator 
    \begin{equation}
    \L_\sscript{SMEFT} \supset -\dfrac{\kappa_u^{\ast p} \kappa_u^{\ast s} \kappa_u^r \kappa_u^t}{64\pi^2 M^2_{R_u}} (\overline{u}^p \gamma_\mu u^r) (\overline{u}^s \gamma_\mu u^t)
    \end{equation}
at one-loop order. This gives a NP contribution to $ D- \overline{D}$ mixing, which is strongly constrained, especially for the CP-violating contributions. We find that the constraints on the LQ parameters are~\cite{Silvestrini:2018dos}
    \begin{align}
    \big|\Re(\Im)\! \big[ (\kappa_u^{2\ast} \kappa_u^{1})^2\big] \big| \left( \dfrac{\SI{50}{TeV}}{M_{R_u}} \right)^{\!\! 2} &< 0.4 \,(0.013). 
    \end{align}
Albeit that the $ D- \overline{D}$ mixing bounds are weaker than those obtained from the dipole operators, they constrain different components of the $ \kappa_u $ coupling vector.

\subsubsection{Scenario II: The phenomenology of a light $S$} 
\label{sec:SPheno}

In an alternative scenario where $M_S$ is light and $v_\Phi \lesssim M_{R_{d},R_{u}} \lesssim M_{Q,L}$, we encounter a profoundly distinct phenomenology: The interactions of $S$ with SM fermions arise primarily through the mediation of heavy fields, either leptoquarks or vector-like fermions. These interactions can be described by dimension-5 operators such as $\frac{\lambda^*_{u} \kappa^{p}_e}{M^2_{R_u}} \overline q_3 S \Phi^\dagger H e^p$ and $\frac{y_e^p z_q}{M_L} \overline{q}_\alpha S^\alpha H e^p$. As a benchmark, we take one component of the flavor doublet to have mass $M_S \simeq 1$\,TeV, while the other component is heavy. Given the limit on $v_\Phi$ from Section~\ref{sec:ZpPheno}, the couplings of $S$ with SM quarks and leptons are expected to be $\max \! \big\{v_\Phi  v_\sscript{EW} / M^2_{R_{d,u}}, v_\sscript{EW} / M_{Q,L} \big\} \lesssim \mathcal{O}(10^{-4})$. As a result, the flavor bounds due to $S$ exchange analogous to those discussed in the previous section are easily satisfied.

Conversely, a light leptoquark can be directly produced in pairs at the LHC through QCD interactions, $g g \to S S^\dagger$. The overall decay width is parameterized by:
\begin{equation}\label{eq:llpdec}
    \frac{\Gamma_S}{M_S} \approx \frac{1}{16 \pi} \frac{v_\sscript{EW}^2}{\Lambda^2}~.
\end{equation}
The specifics hinge on the couplings and the spectrum. In the following, we take $v_\Phi \leq \Lambda \leq M_{Q,L}$ to demonstrate the potential range, considering the uncertainty in $M_{R_{d},R_{u}}$ and marginal couplings $\mathcal{O}(0.3)$. Consequently, the $S$ leptoquark has a longer lifetime than those typically considered in the standard leptoquark searches. Given that $\tau_S \equiv \Gamma^{-1}_S \gg \Lambda^{-1}_{{\rm QCD}}$, this leptoquark undergoes hadronization into a neutral or charged spin-$1/2$ hadron before its decay. Its phenomenology at colliders as a function of $v_\Phi$ provides valuable insights about the scale of the flavor breaking. The search strategies vary based on the lifetime~\cite{Lee:2018pag, Calibbi:2021fld}:
\begin{itemize}
    
    \item {\bf Prompt decay} occurs when $c \tau_S \lesssim 1 \,\mu$m. Conventional leptoquark searches~\cite{CMS:2018oaj, CMS:2018iye} remain relevant (assuming hadronization effects can be ignored). When the decay width is dominated by $R_{d}/R_u$ exchange, the main decay modes involve a combination of third and lighter generations.
    
    \item {\bf Displaced decay} occurs when $1\,\mu\text{m} \ll c \tau_S \lesssim 10$\,m. A dedicated search strategy should follow in the steps of Ref.~\cite{ATLAS:2020xyo}.
    
    \item {\bf Detector-stable decay} occurs when $c \tau_S \gtrsim 10$\,m. The signature involves exotic ionizing tracks suitable for $\frac{d E}{d x}$ searches~\cite{ATLAS:2022pib}.

    \item {\bf Cosmology}: When $v_\Phi$ is exceptionally large, the  longevity of $ S $ can influence cosmological evolution. However, if LQ decays before the onset of Big Bang Nucleosynthesis (BBN), these bounds are satisfied. For a more careful treatment see, e.g.,~\cite{Gross:2018zha, Kawasaki:2017bqm}.    
\end{itemize}

\begin{figure}[t]
         \centering
         \includegraphics[width=0.5\textwidth]{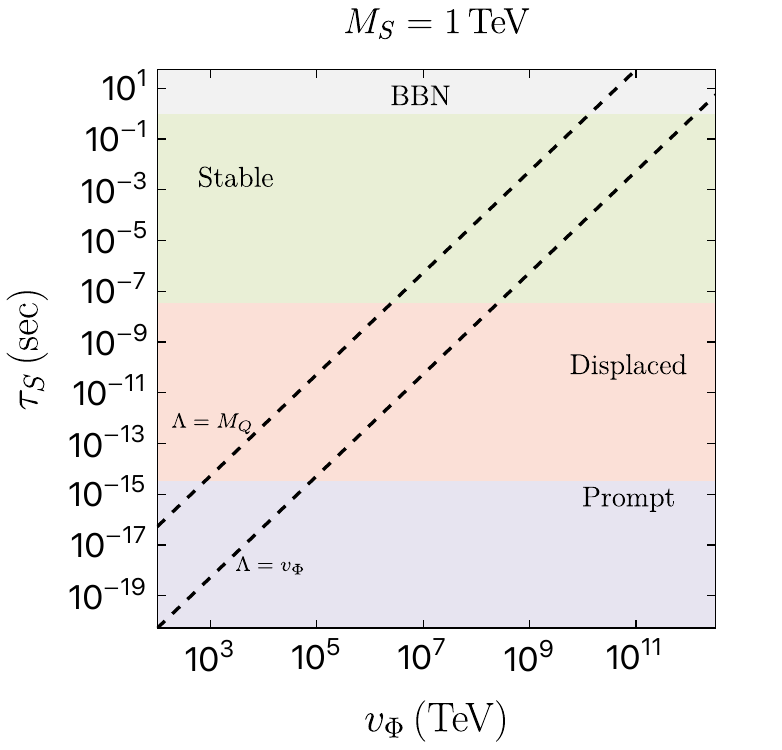}
        \caption{Proper lifetime of the $S$ leptoquark in seconds for $M_S = 1$\,TeV as a function of the \gf breaking scale. See Section~\ref{sec:SPheno} for details.}
        \label{fig:LLP}
\end{figure}

Fig.~\ref{fig:LLP} displays the proper lifetime from Eq.~\eqref{eq:llpdec} as a function of the VEV $v_\Phi$ for $M_S \simeq 1$\,TeV. The two predictions correspond to $\Lambda = v_\Phi$ and $\Lambda = M_{Q}\equiv 100 \,v_\Phi$, respectively. The colored areas in the figure roughly indicate where one of the aforementioned search strategies becomes relevant, serving as a visual guide for the reader. Remarkably, the discovery of a leptoquark via a distinct signature could indirectly shed light on the scale of flavor dynamics in the deep UV. These searches will greatly benefit from the increase of collider energy at the future hadron collider (FCC-hh)~\cite{FCC:2018vvp, MoEDAL-MAPP:2022kyr}.

\section{Discussion and outlook} 
\label{sec:conc}
In this paper, we have outlined a minimal version of a flavor model based on the horizontal $ \SU(2)_{q+\ell} $ gauge group for the light generations. We have demonstrated a very rich phenomenology and upcoming experimental results can potentially discover various low-energy signs of the model. While it is UV complete in its own right, there remain several open questions and interesting possibilities to explore in future work.

\paragraph{Neutrino masses}
So far, our primary focus has been on flavor hierarchies within the charged fermion sector. Let us now delve deeper into the realization of neutrino masses and mixings consistent with the neutrino oscillations data (see, e.g.,~\cite{Esteban:2020cvm}). A straightforward mechanism for generating small neutrino masses is provided by the standard type-I seesaw mechanism~\cite{Minkowski:1977sc, Yanagida:1980xy, Gell-Mann:1979vob}. Let us add three copies of right-handed neutrinos $\nu_p$, where $p$ is the flavor index. The new fields are singlets under the $ G_{\sscript{SM}} \times \SU(2)_{q+\ell}$ gauge symmetry, allowing for a lepton number--violating Majorana mass term $\mathcal{L} \supset - \frac{1}{2} M^{pr}_\RR \nu_p \nu_r$, where $M^{pr}_\RR$ is an arbitrary complex-symmetric mass matrix. 

A scenario with very large Majorana masses (in comparison to the EW scale) explains the smallness of the active neutrino masses,
\begin{equation}\label{eq:NeutrinoMasses}
    m_{\nu_\LL} \simeq - M_D M^{-1}_\RR M^{\intercal}_D \simeq U^{\intercal} \widehat m_{\nu_\LL} U\,,
\end{equation}
via the high-scale see-saw mechanism. Here $M_D \equiv Y_\nu v_{\sscript{EW}}$ is the Dirac mass matrix, $U$ the unitary PMNS mixing matrix, and $\widehat m_{\nu_L}$ a diagonal, active neutrino mass matrix.
This formula disregards a contribution to the PMNS mixing from the (perturbative) charged lepton mixing matrix $L'_e$.\footnote{$U$ does not depend on the potentially large rotation $V_\ell$, which is common to both charged and neutral leptons.} Without additional ingredients, our model predicts $ Y_\nu $ to have a similar hierarchical structure to the charged lepton Yukawa matrix~\eqref{eq:ElectronYuk}. 
When confronted with the observed neutrino parameters, especially the large PMNS mixing angles, a hierarchical $ M_D $ results in a hierarchical Majorana mass matrix $M_\RR$ per Eq.~\eqref{eq:NeutrinoMasses}. 
In the spirit of this paper, such a hierarchy craves an explanation. 

A possible resolution comes from a mechanism to generate anarchic $Y_\nu$. To this end, we can extend the field content with a single vector-like fermion representation $N_{\LL,\RR}\sim (\rep{ 1},\, \rep{ 1},\, 0,\, \rep{ 2})$. When the mass of this field is comparable to $v_\Phi$, marginal interactions $\bar \ell_\alpha \widetilde H N^\alpha$ and $\overline{N}_\alpha \parwidetilde{\Phi}^\alpha \nu_p$ wash out the hierarchy in $Y_\nu$. In this case, the required Majorana mass matrix $M_\RR$ is also anarchic. This is an elegant solution, provided one accepts the coincidence of scales $M_N \sim v_{\Phi}$.

\paragraph{Higgs hierachy problem}
Another obvious limitation of our model is the little Higgs hierarchy problem. One reason for exploring a low-scale resolution to the flavor puzzle is to circumvent potential destabilizing effects on the Higgs potential. However, the flavor constraints from Section~\ref{sec:ZpPheno} already require $v_\Phi \gtrsim \SI{300}{TeV}$, and,  thus, $M_{Q,L} \gtrsim \SI{30}{PeV}$, meaning that the one-loop corrections due to Eq.~\eqref{eq:4} introduce corrections to $\mu^2_H $ of order $M_{Q,L}^2/(16\pi^2)$. It is worth noting that in this model, the Higgs mass is not calculable. But if the model were to be supersymmetrized above the VLF scale, this estimation would accurately reflect the fine-tuning. A possible way forward is to add extra light fields to tame the radiative corrections or implement the mechanism starting from the MSSM.

\paragraph{Embedding in the Pati--Salam gauge group}
The construction of the loops giving mass to the first generation fermions (Fig.~\ref{fig:loop-level_masses}) is very suggestive of quark--lepton unification; compared to the tree-level dimension-5 operators, the loops exchange the vector-like quark for a lepton and vice-versa.   
To this end, a Pati--Salam (PS)~\cite{Pati:1974yy} scenario seems like the way forward. We imagine that our model gets embedded in a $ \SU(4) \times \SU(2)_\LL \times \SU(2)_\RR \times \SU(2)_{q+\ell} $ gauge group in the UV. 

At a simplified level, one can imagine starting with the simpler $ \SU(4) \times \SU(2)_\LL \times \U(1)_\RR \times \SU(2)_{q+\ell} $ group, having traded right-handed isospin for a $ \U(1)_\RR $ charge~\cite{FileviezPerez:2013zmv, FileviezPerez:2023umx}. Remarkably, all the five scalar fields of our model fit into just two irreducible scalar representations of the new gauge group, namely $ H, R_u, R_d \subseteq \Sigma_H \sim (\rep{15},\, \rep{2},\, 1/2,\, \rep{1}) $ and $ \Phi, S \subseteq \Sigma_\Phi \sim (\rep{15},\, \rep{1},\, 0,\, \rep{2}) $. The fermions follow the usual embedding along the PS unification: all the fermions can be arranged in chiral fields $ \psi_\LL \sim (\rep{4},\, \rep{2},\, 0,\, \rep{2}) $, $ \psi^3_\LL \sim (\rep{4},\, \rep{2},\, 0,\, \rep{1}) $, $ \psi^p_{u,\RR} \sim (\rep{4},\, \rep{1},\, 1/2,\, \rep{1}) $, and $ \psi^p_{d,\RR} \sim (\rep{4},\, \rep{1},\, \eminus 1/2,\, \rep{1}) $; along with a vector-like fermion $  \Psi_{\LL,\RR} \sim (\rep{4},\, \rep{2},\, 0,\, \rep{1}) $. This embedding also accounts for the introduction of three right-handed neutrinos. We leave a detailed analysis of this model for future work.

On top of the usual benefits of quark--lepton unification, the embedding into $ \SU(4) $ resolves one of the open problems of our model; it accounts for the degeneracy $ M_Q \sim M_\LL $ as required for a satisfactory explanation of the SM flavor hierarchies. Without such a mechanism, there is no a priori reason for the two fundamental mass parameters to coincide. Thus, with the quark--lepton unification, the model accounts for the six charged fermion mass hierarchies on the basis of a single hierarchy of scales in the UV. The hierarchical CKM matrix comes along for free.

\paragraph{Cosmological imprints}
Another exciting avenue to explore includes potential cosmological imprints of the model. The \gf spontaneous symmetry breaking will, under the right conditions, give rise to a strong first-order phase transition in the early universe, which would produce a stochastic gravitational wave background. The low-scale breaking consistent with the current flavor physics constraints, $v_\Phi \sim 10^3$\,TeV, would be ideal for the future ground-based gravitational wave observatories such as the Einstein Telescope~\cite{Maggiore:2019uih} and the Cosmic Explorer~\cite{Evans:2021gyd}. Discovering such a spectacular signal, together with the predicted pattern of deviations in precision flavor physics, would represent a significant step towards uncovering flavor dynamics in the UV, similarly to~\cite{Greljo:2019xan}.

\section*{Acknowledgments}

We thank Svjetlana Fajer, Peter Stangl, and Stefan Antusch for the useful discussion. This work has received funding from the Swiss National Science Foundation (SNF) through the Eccellenza Professorial Fellowship ``Flavor Physics at the High Energy Frontier'' project number 186866.

\appendix 
\renewcommand{\thesection}{\Alph{section}}
\renewcommand{\thesubsection}{\Alph{section}.\arabic{subsection}}
\setcounter{section}{0}

\section{Perturbative diagonalization of the Yukawa matrices} \label{app:perturbative_diagonalization}

The hierarchical structure of the Higgs coupling matrices to SM fermions generated by the UV model, Eq.~\eqref{eq:SM_Yukawas}, combined with the suitable parametrization of the UV couplings in Eqs.~(\ref{eq:coupling_parametrization_1}, \ref{eq:coupling_parametrization_2}, \ref{eq:coupling_parametrization_3})  allow for diagonalization with perturbative methods. We, therefore, obtain good analytic approximations of both the Higgs couplings to the mass eigenstates and the rotation matrices rotating the fields between the gauge and mass eigenstates, defined in Eq.~\eqref{eq:Yukawa_SVD}. We report only the leading contribution to every individual entry in the matrices.

\subsection{Quark sector}

With the chosen parametrization of the UV couplings, the singular value of the Higgs Yukawa couplings (the couplings of the Higgs boson to the fermion mass eigenstates) are  
    \begin{equation} \label{eq:mass_eigenvalues_quarks}
    \begin{split}
    \widehat{Y}_u &\simeq \diag \! \big(b_q \tilde{\lambda}_{u} z_{u1}, \, a_q y_{u2},\, x_{u3} \big), \\ 
    \widehat{Y}_d &\simeq \diag \! \big(b_q \tilde{\lambda}_{d} z_{d1}, \, a_q y_{d2},\, x_{d3} \big).
    \end{split}
    \end{equation}
Matching these to the SM values provides a quick fix to several of the UV Yukawa couplings. The corresponding rotation matrices are 
    \begin{align}
    L_u &\simeq \begin{pmatrix}
        1 & \tfrac{b_q \tilde{\lambda}_{u} z_{u2}}{a_q y_{u2}} & \tfrac{b_q \tilde{\lambda}_{u} z_{u3}}{ x_{u3}} \\
        \eminus \tfrac{b_q \tilde{\lambda}_{u} z_{u2}^\ast}{a_q y_{u2}} & 1 &  \tfrac{a_q  y_{u3}}{x_{u3}}\\
        \tfrac{b_q \tilde{\lambda}_{u}}{x_{u3}} \! \big[ \tfrac{y_{u3}^\ast z_{u2}^\ast}{y_{u2}} -   z_{u3}^\ast \big]	& \eminus  \tfrac{a_q y_{u3}^\ast }{x_{u3}}& 1
        \end{pmatrix},\\
    R_u &\simeq \begin{pmatrix}
        1 & \tfrac{b_q \lambda^\ast_{u} z_{u1}}{a_q y_{u2}} & \tfrac{a_q b_q \lambda^\ast_{u} y_{u3} z_{u1}}{ x^2_{u3}} \\
        \eminus \tfrac{b_q \lambda_{u} z_{u1}}{a_q y_{u2}} & 1 &  \tfrac{a_q^2  y_{u2} y_{u3}}{x_{u3}^2}\\
        \tfrac{b_q^2 \tilde{\lambda}_{u}^2 z_{u1}}{x^2_{u3}} \! \big[ \tfrac{y_{u3}^\ast z_{u2}^\ast}{y_{u2}} - z_{u3}^\ast \big]	& \eminus  \tfrac{a_q^2 y_{u2} y_{u3}^\ast }{x_{u3}^2}& 1
        \end{pmatrix}, \nonumber
    \end{align}
for the up-type quarks and 
    \begin{align}
    L_d &\simeq \begin{pmatrix}
        1 & \tfrac{b_q \tilde{\lambda}_{d} z_{d2}}{a_q y_{d2}} & \tfrac{b_q \tilde{\lambda}_{d} z_{d3}}{ x_{d3}} \\
        \eminus \tfrac{b_q \tilde{\lambda}_{d} z_{d2}^\ast}{a_q y_{d2}} & 1 &  \tfrac{a_q  y_{d3}}{x_{d3}}\\
        \tfrac{b_q \tilde{\lambda}_{d}}{x_{d3}} \!\big[ \tfrac{y_{d3} z_{d2}^\ast}{y_{d2}} -   z_{d3}^\ast \big]	& \eminus  \tfrac{a_q y_{d3} }{x_{d3}}& 1
        \end{pmatrix},\\
    R_d &\simeq \begin{pmatrix}
        1 & \tfrac{b_q \lambda^\ast_{d} z_{d1}}{a_q y_{d2}} & \tfrac{a_q b_q \lambda^\ast_{d} y_{d3} z_{d1}}{ x^2_{d3}} \\
        \eminus \tfrac{b_q \lambda_{d} z_{d1}}{a_q y_{d2}} & 1 &  \tfrac{a_q^2  y_{d2} y_{d3}}{x_{d3}^2}\\
        \tfrac{b_q^2 \tilde{\lambda}_{d}^2 z_{d1}}{x^2_{d3}} \! \big[ \tfrac{y_{d3} z_{d2}^\ast}{y_{d2}} - z_{d3}^\ast \big]	& \eminus  \tfrac{a_q^2 y_{d2} y_{d3}}{x_{d3}^2}& 1
        \end{pmatrix}, \nonumber
    \end{align}
for the down-type quarks.

\subsection{Lepton sector}
An $ \mathcal{O}(1) $ rotation is required to diagonalize the lepton Yukawa matrix with the tree-level dimension-5 contribution. It is factored out with the rotation matrix 
    \begin{equation} \label{eq:def_Vl}
    V_\ell = \begin{pmatrix}
        c_{\ell} & s_{\ell} & 0 \\ -s_{\ell} & c_{\ell} &0 \\ 0 &0 &1
        \end{pmatrix}, \qquad s_\ell = \dfrac{\tilde{y}_\ell}{\sqrt{y_\ell^2 + \tilde{y}_\ell^2}}~,
    \end{equation}
where $ s_\ell (c_\ell) = \sin (\cos) \theta_\ell $ for some angle $ \theta_\ell $.

After this rotation, it is possible to do the remaining singular value decomposition with perturbative rotation matrices again. We are let to define the quantities
    \begin{equation}
    A_\ell = \sqrt{y_\ell^2 + \tilde{y}_\ell^2} \dfrac{v_\Phi}{M_L}~, \qquad 
    \begin{pmatrix}
    \tilde{\lambda}_e \\ \lambda_e 
    \end{pmatrix} = 
    \begin{pmatrix}
    c_\ell & -s_\ell \\ s_\ell & c_\ell
    \end{pmatrix}
    \begin{pmatrix}
    \lambda^\ast_u \\ -\tilde{\lambda}^\ast_u 
    \end{pmatrix},
    \end{equation}
and find that the diagonalized lepton Yukawa couplings for the charged leptons are 
    \begin{equation} \label{eq:mass_eigenvalues_leptons}
    \widehat{Y}_e \simeq \diag \! \big(b_\ell \tilde{\lambda}_{e} z_{e1}, \, A_\ell y_{e2},\, x_{e3} \big)~.
    \end{equation}
The left-handed rotation matrix is given by $ L_e = V_\ell L_e' $, and we have  
    \begin{align}
    L_e' &\simeq \begin{pmatrix}
        1 & \tfrac{b_\ell \tilde{\lambda}_{e} z_{e2}}{A_\ell y_{e2}} & \tfrac{b_\ell \tilde{\lambda}_{e} z_{e3}}{ x_{e3}} \\
        \eminus \tfrac{b_\ell \tilde{\lambda}^\ast_{e} z_{e2}^\ast}{A_\ell y_{e2}} & 1 &  \tfrac{A_\ell  y_{e3}}{x_{e3}}\\
        \tfrac{b_\ell \tilde{\lambda}^\ast_{e}}{x_{e3}} \! \big[ \tfrac{y_{e3} z_{e2}^\ast}{y_{e2}} -   z_{e3}^\ast \big]	& \eminus  \tfrac{A_\ell y_{e3} }{x_{e3}}& 1
        \end{pmatrix}, \\
    R_e & \simeq \begin{pmatrix}
        1 & \tfrac{b_\ell \lambda^\ast_{e} z_{e1}}{A_\ell y_{e2}} & \tfrac{A_\ell b_\ell \lambda^\ast_{e} y_{e3} z_{e1}}{ x^2_{e3}} \\
        \eminus \tfrac{b_\ell \lambda_{e} z_{e1}}{A_\ell y_{e2}} & 1 &  \tfrac{A_\ell^2  y_{e2} y_{e3}}{x^2_{e3}}\\
        \tfrac{b^2_\ell |\tilde{\lambda}_{e}|^2 z_{e1}}{x^2_{e3}} \! \big[ \tfrac{y_{e3} z_{e2}^\ast}{y_{e2}} -   z_{e3}^\ast \big]	& \eminus  \tfrac{A^2_\ell y_{e2} y_{e3} }{x^2_{e3}}& 1
        \end{pmatrix}. \nonumber
    \end{align}
On the background of the perturbative formulas for the Higgs couplings to Higgs couplings and fermion mixing matrices, we can establish a benchmark point for the UV parameters that faithfully reproduce the experimental measurements.

\subsection{Benchmark values}
To illustrate some realistic parameters for the model, we consider here a \emph{single} benchmark point. We take the breaking scale of \gf to be fairly small at $ v_\Phi = \SI{e3}{TeV} $. The LQ masses influence the Higgs Yukawa couplings only by controlling the RG running. Thus, their exact value is less important and we simply set the renormalization scale $ \mu = \SI{e3}{TeV} $ as roughly the scale of the LQs. 

We match the model to the SM Yukawas at the scale $ \mu $ and so take into account the running of the SM parameter, taken from Ref.~\cite{Martin:2019lqd}, using the RG functions of \texttt{RGBeta}~\cite{Thomsen:2021ncy} at loop order 3--2--2 (gauge--Yukawa--quartic). The target values of the Higgs couplings to the fermion mass eigenstates at $ \mu = \SI{e3}{TeV} $ are 
    \begin{align}
    \big(y_u,\, y_c,\, y_t \big)_\sscript{SM} &= \big(\num{4.54e-6},\,\num{2.29e-3},\, \num{0.667}\big), \nonumber\\
    \big(y_d,\, y_s,\, y_b \big)_\sscript{SM} &= \big(\num{9.95e-6},\,\num{1.98e-4},\, \num{0.0100}\big),\nonumber\\
    \big(y_e,\, y_\mu,\, y_\tau \big)_\sscript{SM} &= \big(\num{2.87e-6},\,\num{6.05e-4},\, \num{0.0103}\big). 
    \end{align}
To a good approximation, the CKM matrix of the SM does not run, and we use the PDG values~\cite{Workman:2022ynf} directly at the scale $ \mu $.

We find that VLF masses $ M_Q = M_L = 100\, v_\Phi $ are a good starting point for producing the hierarchy of the SM Higgs couplings. We let 
    \begin{equation}
    \begin{gathered}
        y_q= 0.25~, \quad z_q = 0.3~, \quad y_\ell = 0.2~, \quad \tilde{y}_\ell = 0~, \\
        z_\ell = 0.1~, \quad \tilde{\lambda}_u= \tilde{\lambda}_d = 0.3~,\quad \lambda_u = 0.2~.
    \end{gathered}
    \end{equation}
such that 
    \begin{align}
    a_q &= \num{2.5e-3}, & b_q & = \num{1.6e-4}, \\
    A_\ell &= \num{2e-3}, & b_\ell & = \num{1.6e-4}.
    \end{align}
With these choices, the SM values for the Yukawa parameters allow us to fix the diagonal values of the vectors from comparison with Eqs.~(\ref{eq:mass_eigenvalues_quarks}, \ref{eq:mass_eigenvalues_leptons}): 
    \begin{align}
    \big(z_{u1},\, y_{u2},\, x_{u3}\big) &= \big(0.097,\, 0.91,\, 0.67 \big), \nonumber \\
    \big(z_{d1},\, y_{d2},\, x_{d3}\big) &= \big(0.21,\, 0.079,\, 0.010 \big), \\
    \big(z_{e1},\, y_{e2},\, x_{e3}\big) &= \big(0.092,\, 0.30,\, 0.010\big). \nonumber
    \end{align}

The SM mixing angles, captured by the CKM matrix, provide only limited information to fix the remaining parameters in $ y_f^p $ and $ z_f^p $. 
The compressed mass hierarchy in the down quark sector as compared to the up quark sector, ensures that the mixing contributions from off-diagonal down-type Yukawa couplings are enhanced.  
Hence, we take the CKM matrix~\eqref{eq:ckm_approx} to be exclusively due to down-type contributions in this benchmark. To a good approximation, the SM values for the CKM are obtained with
    \begin{align}
    y_{d3} &= 0.16\,, & z_{d2} &= 0.95e^{i \alpha} \,, & z_{d3} &= 0.77e^{i (\alpha-1.20)}\,,
    \end{align}
for any phase $ \alpha $. Non-zero up-type contributions to the CKM matrix will shift these values only by small amounts.

\bibliographystyle{JHEP}
\bibliography{refs.bib}

\providecommand{\href}[2]{#2}\begingroup\raggedright\begin{thebibliography}{100}

\bibitem{tHooft:1979rat}
G.~'t~Hooft, {\it {Naturalness, chiral symmetry, and spontaneous chiral
  symmetry breaking}},  {\em NATO Sci. Ser. B} {\bf 59} (1980) 135--157.

\bibitem{Froggatt:1978nt}
C.~D. Froggatt and H.~B. Nielsen, {\it {Hierarchy of Quark Masses, Cabibbo
  Angles and CP Violation}},  {\em Nucl. Phys. B} {\bf 147} (1979) 277--298.

\bibitem{Leurer:1992wg}
M.~Leurer, Y.~Nir, and N.~Seiberg, {\it {Mass matrix models}},  {\em Nucl.
  Phys. B} {\bf 398} (1993) 319--342,
  [\href{http://arxiv.org/abs/hep-ph/9212278}{{\tt hep-ph/9212278}}].

\bibitem{Leurer:1993gy}
M.~Leurer, Y.~Nir, and N.~Seiberg, {\it {Mass matrix models: The Sequel}},
  {\em Nucl. Phys. B} {\bf 420} (1994) 468--504,
  [\href{http://arxiv.org/abs/hep-ph/9310320}{{\tt hep-ph/9310320}}].

\bibitem{Fedele:2020fvh}
M.~Fedele, A.~Mastroddi, and M.~Valli, {\it {Minimal Froggatt-Nielsen
  textures}},  {\em JHEP} {\bf 03} (2021) 135,
  [\href{http://arxiv.org/abs/2009.05587}{{\tt arXiv:2009.05587}}].

\bibitem{Cornella:2023zme}
C.~Cornella, D.~Curtin, E.~T. Neil, and J.~O. Thompson, {\it {Mapping and
  Probing Froggatt-Nielsen Solutions to the Quark Flavor Puzzle}},
  \href{http://arxiv.org/abs/2306.08026}{{\tt arXiv:2306.08026}}.

\bibitem{Asadi:2023ucx}
P.~Asadi, A.~Bhattacharya, K.~Fraser, S.~Homiller, and A.~Parikh, {\it
  {Wrinkles in the Froggatt-Nielsen Mechanism and Flavorful New Physics}},
  \href{http://arxiv.org/abs/2308.01340}{{\tt arXiv:2308.01340}}.

\bibitem{Smolkovic:2019jow}
A.~Smolkovi\v{c}, M.~Tammaro, and J.~Zupan, {\it {Anomaly free Froggatt-Nielsen
  models of flavor}},  {\em JHEP} {\bf 10} (2019) 188,
  [\href{http://arxiv.org/abs/1907.10063}{{\tt arXiv:1907.10063}}]. [Erratum:
  JHEP 02, 033 (2022)].

\bibitem{Grinstein:2010ve}
B.~Grinstein, M.~Redi, and G.~Villadoro, {\it {Low Scale Flavor Gauge
  Symmetries}},  {\em JHEP} {\bf 11} (2010) 067,
  [\href{http://arxiv.org/abs/1009.2049}{{\tt arXiv:1009.2049}}].

\bibitem{DAgnolo:2012ulg}
R.~T. D'Agnolo and D.~M. Straub, {\it {Gauged flavour symmetry for the light
  generations}},  {\em JHEP} {\bf 05} (2012) 034,
  [\href{http://arxiv.org/abs/1202.4759}{{\tt arXiv:1202.4759}}].

\bibitem{Kaplan:1993ej}
D.~B. Kaplan and M.~Schmaltz, {\it {Flavor unification and discrete nonAbelian
  symmetries}},  {\em Phys. Rev. D} {\bf 49} (1994) 3741--3750,
  [\href{http://arxiv.org/abs/hep-ph/9311281}{{\tt hep-ph/9311281}}].

\bibitem{King:2003rf}
S.~F. King and G.~G. Ross, {\it {Fermion masses and mixing angles from SU (3)
  family symmetry and unification}},  {\em Phys. Lett. B} {\bf 574} (2003)
  239--252, [\href{http://arxiv.org/abs/hep-ph/0307190}{{\tt hep-ph/0307190}}].

\bibitem{Antusch:2008jf}
S.~Antusch, S.~F. King, M.~Malinsky, and G.~G. Ross, {\it {Solving the SUSY
  Flavour and CP Problems with Non-Abelian Family Symmetry and Supergravity}},
  {\em Phys. Lett. B} {\bf 670} (2009) 383--389,
  [\href{http://arxiv.org/abs/0807.5047}{{\tt arXiv:0807.5047}}].

\bibitem{Antusch:2007re}
S.~Antusch, S.~F. King, and M.~Malinsky, {\it {Solving the SUSY Flavour and CP
  Problems with SU(3) Family Symmetry}},  {\em JHEP} {\bf 06} (2008) 068,
  [\href{http://arxiv.org/abs/0708.1282}{{\tt arXiv:0708.1282}}].

\bibitem{Bishara:2015mha}
F.~Bishara, A.~Greljo, J.~F. Kamenik, E.~Stamou, and J.~Zupan, {\it {Dark
  Matter and Gauged Flavor Symmetries}},  {\em JHEP} {\bf 12} (2015) 130,
  [\href{http://arxiv.org/abs/1505.03862}{{\tt arXiv:1505.03862}}].

\bibitem{Linster:2018avp}
M.~Linster and R.~Ziegler, {\it {A Realistic $U(2)$ Model of Flavor}},  {\em
  JHEP} {\bf 08} (2018) 058, [\href{http://arxiv.org/abs/1805.07341}{{\tt
  arXiv:1805.07341}}].

\bibitem{Davighi:2022fer}
J.~Davighi and J.~Tooby-Smith, {\it {Electroweak flavour unification}},  {\em
  JHEP} {\bf 09} (2022) 193, [\href{http://arxiv.org/abs/2201.07245}{{\tt
  arXiv:2201.07245}}].

\bibitem{Barr:1990td}
S.~M. Barr, {\it {A Predictive hierarchical mode of quark and lepton masses}},
  {\em Phys. Rev. D} {\bf 42} (1990) 3150--3159.

\bibitem{Barr:2001vj}
S.~M. Barr, {\it {Flavor without flavor symmetry}},  {\em Phys. Rev. D} {\bf
  65} (2002) 096012, [\href{http://arxiv.org/abs/hep-ph/0106241}{{\tt
  hep-ph/0106241}}].

\bibitem{Ferretti:2006df}
L.~Ferretti, S.~F. King, and A.~Romanino, {\it {Flavour from accidental
  symmetries}},  {\em JHEP} {\bf 11} (2006) 078,
  [\href{http://arxiv.org/abs/hep-ph/0609047}{{\tt hep-ph/0609047}}].

\bibitem{Weinberg:1972ws}
S.~Weinberg, {\it {Electromagnetic and weak masses}},  {\em Phys. Rev. Lett.}
  {\bf 29} (1972) 388--392.

\bibitem{Arkani-Hamed:1996kxn}
N.~Arkani-Hamed, H.-C. Cheng, and L.~J. Hall, {\it {A Supersymmetric theory of
  flavor with radiative fermion masses}},  {\em Phys. Rev. D} {\bf 54} (1996)
  2242--2260, [\href{http://arxiv.org/abs/hep-ph/9601262}{{\tt
  hep-ph/9601262}}].

\bibitem{Altmannshofer:2014qha}
W.~Altmannshofer, C.~Frugiuele, and R.~Harnik, {\it {Fermion Hierarchy from
  Sfermion Anarchy}},  {\em JHEP} {\bf 12} (2014) 180,
  [\href{http://arxiv.org/abs/1409.2522}{{\tt arXiv:1409.2522}}].

\bibitem{Weinberg:2020zba}
S.~Weinberg, {\it {Models of Lepton and Quark Masses}},  {\em Phys. Rev. D}
  {\bf 101} (2020), no.~3 035020, [\href{http://arxiv.org/abs/2001.06582}{{\tt
  arXiv:2001.06582}}].

\bibitem{Baker:2020vkh}
M.~J. Baker, P.~Cox, and R.~R. Volkas, {\it {Has the Origin of the Third-Family
  Fermion Masses been Determined?}},  {\em JHEP} {\bf 04} (2021) 151,
  [\href{http://arxiv.org/abs/2012.10458}{{\tt arXiv:2012.10458}}].

\bibitem{Porto:2007ed}
R.~A. Porto and A.~Zee, {\it {The Private Higgs}},  {\em Phys. Lett. B} {\bf
  666} (2008) 491--495, [\href{http://arxiv.org/abs/0712.0448}{{\tt
  arXiv:0712.0448}}].

\bibitem{Hill:2019ldq}
C.~T. Hill, P.~A.~N. Machado, A.~E. Thomsen, and J.~Turner, {\it {Scalar
  Democracy}},  {\em Phys. Rev. D} {\bf 100} (2019), no.~1 015015,
  [\href{http://arxiv.org/abs/1902.07214}{{\tt arXiv:1902.07214}}].

\bibitem{Baek:2023cfy}
S.~Baek, J.~Kersten, P.~Ko, and L.~Velasco-Sevilla, {\it {An Unfamiliar Way to
  Generate the Hierarchy of Standard Model Fermion Masses}},
  \href{http://arxiv.org/abs/2309.07788}{{\tt arXiv:2309.07788}}.

\bibitem{Barbieri:1994cx}
R.~Barbieri, G.~R. Dvali, and A.~Strumia, {\it {Fermion masses and mixings in a
  flavor symmetric GUT}},  {\em Nucl. Phys. B} {\bf 435} (1995) 102--114,
  [\href{http://arxiv.org/abs/hep-ph/9407239}{{\tt hep-ph/9407239}}].

\bibitem{Panico:2016ull}
G.~Panico and A.~Pomarol, {\it {Flavor hierarchies from dynamical scales}},
  {\em JHEP} {\bf 07} (2016) 097, [\href{http://arxiv.org/abs/1603.06609}{{\tt
  arXiv:1603.06609}}].

\bibitem{Bordone:2017bld}
M.~Bordone, C.~Cornella, J.~Fuentes-Martin, and G.~Isidori, {\it {A three-site
  gauge model for flavor hierarchies and flavor anomalies}},  {\em Phys. Lett.
  B} {\bf 779} (2018) 317--323, [\href{http://arxiv.org/abs/1712.01368}{{\tt
  arXiv:1712.01368}}].

\bibitem{Greljo:2018tuh}
A.~Greljo and B.~A. Stefanek, {\it {Third family quark\textendash{}lepton
  unification at the TeV scale}},  {\em Phys. Lett. B} {\bf 782} (2018)
  131--138, [\href{http://arxiv.org/abs/1802.04274}{{\tt arXiv:1802.04274}}].

\bibitem{Allanach:2018lvl}
B.~C. Allanach and J.~Davighi, {\it {Third family hypercharge model for $
  {R}_{K^{\left(\ast \right)}} $ and aspects of the fermion mass problem}},
  {\em JHEP} {\bf 12} (2018) 075, [\href{http://arxiv.org/abs/1809.01158}{{\tt
  arXiv:1809.01158}}].

\bibitem{Greljo:2019xan}
A.~Greljo, T.~Opferkuch, and B.~A. Stefanek, {\it {Gravitational Imprints of
  Flavor Hierarchies}},  {\em Phys. Rev. Lett.} {\bf 124} (2020), no.~17
  171802, [\href{http://arxiv.org/abs/1910.02014}{{\tt arXiv:1910.02014}}].

\bibitem{Allwicher:2020esa}
L.~Allwicher, G.~Isidori, and A.~E. Thomsen, {\it {Stability of the Higgs
  Sector in a Flavor-Inspired Multi-Scale Model}},  {\em JHEP} {\bf 01} (2021)
  191, [\href{http://arxiv.org/abs/2011.01946}{{\tt arXiv:2011.01946}}].

\bibitem{Davighi:2023iks}
J.~Davighi and G.~Isidori, {\it {Non-universal gauge interactions addressing
  the inescapable link between Higgs and flavour}},  {\em JHEP} {\bf 07} (2023)
  147, [\href{http://arxiv.org/abs/2303.01520}{{\tt arXiv:2303.01520}}].

\bibitem{Davighi:2023evx}
J.~Davighi and B.~A. Stefanek, {\it {Deconstructed Hypercharge: A Natural Model
  of Flavour}},  \href{http://arxiv.org/abs/2305.16280}{{\tt
  arXiv:2305.16280}}.

\bibitem{FernandezNavarro:2023rhv}
M.~Fern\'andez~Navarro and S.~F. King, {\it {Tri-hypercharge: a separate gauged
  weak hypercharge for each fermion family as the origin of flavour}},  {\em
  JHEP} {\bf 08} (2023) 020, [\href{http://arxiv.org/abs/2305.07690}{{\tt
  arXiv:2305.07690}}].

\bibitem{Randall:1999ee}
L.~Randall and R.~Sundrum, {\it {A Large mass hierarchy from a small extra
  dimension}},  {\em Phys. Rev. Lett.} {\bf 83} (1999) 3370--3373,
  [\href{http://arxiv.org/abs/hep-ph/9905221}{{\tt hep-ph/9905221}}].

\bibitem{Arkani-Hamed:1999ylh}
N.~Arkani-Hamed and M.~Schmaltz, {\it {Hierarchies without symmetries from
  extra dimensions}},  {\em Phys. Rev. D} {\bf 61} (2000) 033005,
  [\href{http://arxiv.org/abs/hep-ph/9903417}{{\tt hep-ph/9903417}}].

\bibitem{Fuentes-Martin:2022xnb}
J.~Fuentes-Martin, G.~Isidori, J.~M. Lizana, N.~Selimovic, and B.~A. Stefanek,
  {\it {Flavor hierarchies, flavor anomalies, and Higgs mass from a warped
  extra dimension}},  {\em Phys. Lett. B} {\bf 834} (2022) 137382,
  [\href{http://arxiv.org/abs/2203.01952}{{\tt arXiv:2203.01952}}].

\bibitem{Kaplan:1991dc}
D.~B. Kaplan, {\it {Flavor at SSC energies: A New mechanism for dynamically
  generated fermion masses}},  {\em Nucl. Phys. B} {\bf 365} (1991) 259--278.

\bibitem{Ferretti:2014qta}
G.~Ferretti, {\it {UV Completions of Partial Compositeness: The Case for a
  SU(4) Gauge Group}},  {\em JHEP} {\bf 06} (2014) 142,
  [\href{http://arxiv.org/abs/1404.7137}{{\tt arXiv:1404.7137}}].

\bibitem{Sannino:2016sfx}
F.~Sannino, A.~Strumia, A.~Tesi, and E.~Vigiani, {\it {Fundamental partial
  compositeness}},  {\em JHEP} {\bf 11} (2016) 029,
  [\href{http://arxiv.org/abs/1607.01659}{{\tt arXiv:1607.01659}}].

\bibitem{Patel:2017pct}
K.~M. Patel, {\it {Clockwork mechanism for flavor hierarchies}},  {\em Phys.
  Rev. D} {\bf 96} (2017), no.~11 115013,
  [\href{http://arxiv.org/abs/1711.05393}{{\tt arXiv:1711.05393}}].

\bibitem{Alonso:2018bcg}
R.~Alonso, A.~Carmona, B.~M. Dillon, J.~F. Kamenik, J.~Martin~Camalich, and
  J.~Zupan, {\it {A clockwork solution to the flavor puzzle}},  {\em JHEP} {\bf
  10} (2018) 099, [\href{http://arxiv.org/abs/1807.09792}{{\tt
  arXiv:1807.09792}}].

\bibitem{Altmannshofer:2021qwx}
W.~Altmannshofer and S.~A. Gadam, {\it {Supersymmetric flavor clockwork
  model}},  {\em Phys. Rev. D} {\bf 104} (2021), no.~3 035030,
  [\href{http://arxiv.org/abs/2106.09869}{{\tt arXiv:2106.09869}}].

\bibitem{Feruglio:2021dte}
F.~Feruglio, V.~Gherardi, A.~Romanino, and A.~Titov, {\it {Modular invariant
  dynamics and fermion mass hierarchies around $\tau = i$}},  {\em JHEP} {\bf
  05} (2021) 242, [\href{http://arxiv.org/abs/2101.08718}{{\tt
  arXiv:2101.08718}}].

\bibitem{Alonso:2011yg}
R.~Alonso, M.~B. Gavela, L.~Merlo, and S.~Rigolin, {\it {On the scalar
  potential of minimal flavour violation}},  {\em JHEP} {\bf 07} (2011) 012,
  [\href{http://arxiv.org/abs/1103.2915}{{\tt arXiv:1103.2915}}].

\bibitem{DAmbrosio:2002vsn}
G.~D'Ambrosio, G.~F. Giudice, G.~Isidori, and A.~Strumia, {\it {Minimal flavor
  violation: An Effective field theory approach}},  {\em Nucl. Phys. B} {\bf
  645} (2002) 155--187, [\href{http://arxiv.org/abs/hep-ph/0207036}{{\tt
  hep-ph/0207036}}].

\bibitem{Barbieri:1995uv}
R.~Barbieri, G.~R. Dvali, and L.~J. Hall, {\it {Predictions from a U(2) flavor
  symmetry in supersymmetric theories}},  {\em Phys. Lett. B} {\bf 377} (1996)
  76--82, [\href{http://arxiv.org/abs/hep-ph/9512388}{{\tt hep-ph/9512388}}].

\bibitem{Barbieri:2011ci}
R.~Barbieri, G.~Isidori, J.~Jones-Perez, P.~Lodone, and D.~M. Straub, {\it
  {$U(2)$ and Minimal Flavour Violation in Supersymmetry}},  {\em Eur. Phys. J.
  C} {\bf 71} (2011) 1725, [\href{http://arxiv.org/abs/1105.2296}{{\tt
  arXiv:1105.2296}}].

\bibitem{Kagan:2009bn}
A.~L. Kagan, G.~Perez, T.~Volansky, and J.~Zupan, {\it {General Minimal Flavor
  Violation}},  {\em Phys. Rev. D} {\bf 80} (2009) 076002,
  [\href{http://arxiv.org/abs/0903.1794}{{\tt arXiv:0903.1794}}].

\bibitem{Bento:2023owf}
M.~P. Bento, J.~P. Silva, and A.~Trautner, {\it {The Basis Invariant Flavor
  Puzzle}},  \href{http://arxiv.org/abs/2308.00019}{{\tt arXiv:2308.00019}}.

\bibitem{Babu:1990hu}
K.~S. Babu and R.~N. Mohapatra, {\it {Radiative fermion masses and large
  neutrino magnetic moment: A Unified picture}},  {\em Phys. Rev. D} {\bf 43}
  (1991) 2278--2282.

\bibitem{Shaw:1992gk}
D.~S. Shaw and R.~R. Volkas, {\it {Systematic study of fermion masses and
  mixing angles in horizontal SU(2) gauge theory}},  {\em Phys. Rev. D} {\bf
  47} (1993) 241--255, [\href{http://arxiv.org/abs/hep-ph/9211209}{{\tt
  hep-ph/9211209}}].

\bibitem{Darme:2023nsy}
L.~Darm\'e, A.~Deandrea, and F.~Mahmoudi, {\it {Gauge $SU(2)_f$ flavour
  transfers}},  \href{http://arxiv.org/abs/2307.09595}{{\tt arXiv:2307.09595}}.

\bibitem{Arnold:2012sd}
J.~M. Arnold, B.~Fornal, and M.~B. Wise, {\it {Simplified models with baryon
  number violation but no proton decay}},  {\em Phys. Rev. D} {\bf 87} (2013)
  075004, [\href{http://arxiv.org/abs/1212.4556}{{\tt arXiv:1212.4556}}].

\bibitem{Murgui:2021bdy}
C.~Murgui and M.~B. Wise, {\it {Scalar leptoquarks, baryon number violation,
  and Pati-Salam symmetry}},  {\em Phys. Rev. D} {\bf 104} (2021), no.~3
  035017, [\href{http://arxiv.org/abs/2105.14029}{{\tt arXiv:2105.14029}}].

\bibitem{Dorsner:2022twk}
I.~Dor\v{s}ner, S.~Fajfer, and O.~Sumensari, {\it {Triple-leptoquark
  interactions for tree- and loop-level proton decays}},  {\em JHEP} {\bf 05}
  (2022) 183, [\href{http://arxiv.org/abs/2202.08287}{{\tt arXiv:2202.08287}}].

\bibitem{Crivellin:2021ejk}
A.~Crivellin and L.~Schnell, {\it {Complete Lagrangian and set of Feynman rules
  for scalar leptoquarks}},  {\em Comput. Phys. Commun.} {\bf 271} (2022)
  108188, [\href{http://arxiv.org/abs/2105.04844}{{\tt arXiv:2105.04844}}].

\bibitem{Davighi:2020qqa}
J.~Davighi, M.~Kirk, and M.~Nardecchia, {\it {Anomalies and accidental
  symmetries: charging the scalar leptoquark under L$_{\mu}$ \ensuremath{-}
  L$_{\tau}$}},  {\em JHEP} {\bf 12} (2020) 111,
  [\href{http://arxiv.org/abs/2007.15016}{{\tt arXiv:2007.15016}}].

\bibitem{Greljo:2021xmg}
A.~Greljo, P.~Stangl, and A.~E. Thomsen, {\it {A model of muon anomalies}},
  {\em Phys. Lett. B} {\bf 820} (2021) 136554,
  [\href{http://arxiv.org/abs/2103.13991}{{\tt arXiv:2103.13991}}].

\bibitem{Davighi:2022qgb}
J.~Davighi, A.~Greljo, and A.~E. Thomsen, {\it {Leptoquarks with exactly stable
  protons}},  {\em Phys. Lett. B} {\bf 833} (2022) 137310,
  [\href{http://arxiv.org/abs/2202.05275}{{\tt arXiv:2202.05275}}].

\bibitem{Fuentes-Martin:2022jrf}
J.~Fuentes-Mart\'\i{}n, M.~K\"onig, J.~Pag\`es, A.~E. Thomsen, and F.~Wilsch,
  {\it {A proof of concept for matchete: an automated tool for matching
  effective theories}},  {\em Eur. Phys. J. C} {\bf 83} (2023), no.~7 662,
  [\href{http://arxiv.org/abs/2212.04510}{{\tt arXiv:2212.04510}}].

\bibitem{Silvestrini:2018dos}
L.~Silvestrini and M.~Valli, {\it {Model-independent Bounds on the Standard
  Model Effective Theory from Flavour Physics}},  {\em Phys. Lett. B} {\bf 799}
  (2019) 135062, [\href{http://arxiv.org/abs/1812.10913}{{\tt
  arXiv:1812.10913}}].

\bibitem{Jenkins:2017jig}
E.~E. Jenkins, A.~V. Manohar, and P.~Stoffer, {\it {Low-Energy Effective Field
  Theory below the Electroweak Scale: Operators and Matching}},  {\em JHEP}
  {\bf 03} (2018) 016, [\href{http://arxiv.org/abs/1709.04486}{{\tt
  arXiv:1709.04486}}].

\bibitem{Marzocca:2021miv}
D.~Marzocca, S.~Trifinopoulos, and E.~Venturini, {\it {From B-meson anomalies
  to Kaon physics with scalar leptoquarks}},  {\em Eur. Phys. J. C} {\bf 82}
  (2022), no.~4 320, [\href{http://arxiv.org/abs/2106.15630}{{\tt
  arXiv:2106.15630}}].

\bibitem{Angelescu:2020uug}
A.~Angelescu, D.~A. Faroughy, and O.~Sumensari, {\it {Lepton Flavor Violation
  and Dilepton Tails at the LHC}},  {\em Eur. Phys. J. C} {\bf 80} (2020),
  no.~7 641, [\href{http://arxiv.org/abs/2002.05684}{{\tt arXiv:2002.05684}}].

\bibitem{BNL:1998apv}
{\bf BNL} Collaboration, D.~Ambrose et~al., {\it {New limit on muon and
  electron lepton number violation from K0(L) ---\ensuremath{>} mu+- e-+
  decay}},  {\em Phys. Rev. Lett.} {\bf 81} (1998) 5734--5737,
  [\href{http://arxiv.org/abs/hep-ex/9811038}{{\tt hep-ex/9811038}}].

\bibitem{SINDRUMII:2006dvw}
{\bf SINDRUM II} Collaboration, W.~H. Bertl et~al., {\it {A Search for muon to
  electron conversion in muonic gold}},  {\em Eur. Phys. J. C} {\bf 47} (2006)
  337--346.

\bibitem{Bernstein:2019fyh}
{\bf Mu2e} Collaboration, R.~H. Bernstein, {\it {The Mu2e Experiment}},  {\em
  Front. in Phys.} {\bf 7} (2019) 1,
  [\href{http://arxiv.org/abs/1901.11099}{{\tt arXiv:1901.11099}}].

\bibitem{Moritsu:2022lem}
{\bf COMET} Collaboration, M.~Moritsu, {\it {Search for Muon-to-Electron
  Conversion with the COMET Experiment \textdagger{}}},  {\em Universe} {\bf 8}
  (2022), no.~4 196, [\href{http://arxiv.org/abs/2203.06365}{{\tt
  arXiv:2203.06365}}].

\bibitem{Jegerlehner:2009ry}
F.~Jegerlehner and A.~Nyffeler, {\it {The Muon g-2}},  {\em Phys. Rept.} {\bf
  477} (2009) 1--110, [\href{http://arxiv.org/abs/0902.3360}{{\tt
  arXiv:0902.3360}}].

\bibitem{ACME:2018yjb}
{\bf ACME} Collaboration, V.~Andreev et~al., {\it {Improved limit on the
  electric dipole moment of the electron}},  {\em Nature} {\bf 562} (2018),
  no.~7727 355--360.

\bibitem{NA62:2021zjw}
{\bf NA62} Collaboration, E.~Cortina~Gil et~al., {\it {Measurement of the very
  rare K$^{+}$\textrightarrow{}$ {\pi}^{+}\nu \overline{\nu} $ decay}},  {\em
  JHEP} {\bf 06} (2021) 093, [\href{http://arxiv.org/abs/2103.15389}{{\tt
  arXiv:2103.15389}}].

\bibitem{KOTO:2018dsc}
{\bf KOTO} Collaboration, J.~K. Ahn et~al., {\it {Search for the $K_L \!\to\!
  \pi^0 \nu \overline{\nu}$ and $K_L \!\to\! \pi^0 X^0$ decays at the J-PARC
  KOTO experiment}},  {\em Phys. Rev. Lett.} {\bf 122} (2019), no.~2 021802,
  [\href{http://arxiv.org/abs/1810.09655}{{\tt arXiv:1810.09655}}].

\bibitem{Buras:2020xsm}
A.~Buras, {\em {Gauge Theory of Weak Decays}}.
\newblock Cambridge University Press, 6, 2020.

\bibitem{Brod:2021hsj}
J.~Brod, M.~Gorbahn, and E.~Stamou, {\it {Updated Standard Model Prediction for
  $K \to \pi \nu \bar{\nu}$ and $\epsilon_K$}},  {\em PoS} {\bf BEAUTY2020}
  (2021) 056, [\href{http://arxiv.org/abs/2105.02868}{{\tt arXiv:2105.02868}}].

\bibitem{NA62KLEVER:2022nea}
{\bf NA62/KLEVER, US Kaon Interest Group, KOTO, LHCb} Collaboration, {\it
  {Searches for new physics with high-intensity kaon beams}},  in {\em
  {Snowmass 2021}}, 4, 2022.
\newblock \href{http://arxiv.org/abs/2204.13394}{{\tt arXiv:2204.13394}}.

\bibitem{HIKE:2022qra}
{\bf HIKE} Collaboration, E.~Cortina~Gil et~al., {\it {HIKE, High Intensity
  Kaon Experiments at the CERN SPS}: {Letter of Intent}},
  \href{http://arxiv.org/abs/2211.16586}{{\tt arXiv:2211.16586}}.

\bibitem{Belle:2017oht}
{\bf Belle} Collaboration, J.~Grygier et~al., {\it {Search for
  $\boldsymbol{B\to h\nu\bar{\nu}}$ decays with semileptonic tagging at
  Belle}},  {\em Phys. Rev. D} {\bf 96} (2017), no.~9 091101,
  [\href{http://arxiv.org/abs/1702.03224}{{\tt arXiv:1702.03224}}]. [Addendum:
  Phys.Rev.D 97, 099902 (2018)].

\bibitem{Belle-II:2021rof}
{\bf Belle-II} Collaboration, F.~Abudin\'en et~al., {\it {Search for
  B+\textrightarrow{}K+\ensuremath{\nu}\ensuremath{\nu}\textasciimacron{}
  Decays Using an Inclusive Tagging Method at Belle II}},  {\em Phys. Rev.
  Lett.} {\bf 127} (2021), no.~18 181802,
  [\href{http://arxiv.org/abs/2104.12624}{{\tt arXiv:2104.12624}}].

\bibitem{GlazovEPS}
A.~Glazov, {\it Talk at european physical society conference on high energy
  physics 2023, 24.08.2023}, .

\bibitem{Allwicher:2023syp}
L.~Allwicher, D.~Becirevic, G.~Piazza, S.~Rosauro-Alcaraz, and O.~Sumensari,
  {\it {Understanding the first measurement of $\mathcal{B}(B\to K \nu
  \bar{\nu})$}},  \href{http://arxiv.org/abs/2309.02246}{{\tt
  arXiv:2309.02246}}.

\bibitem{Bause:2023mfe}
R.~Bause, H.~Gisbert, and G.~Hiller, {\it {Implications of an enhanced $B \to K
  \nu \bar \nu$ branching ratio}},  \href{http://arxiv.org/abs/2309.00075}{{\tt
  arXiv:2309.00075}}.

\bibitem{Felkl:2023ayn}
T.~Felkl, A.~Giri, R.~Mohanta, and M.~A. Schmidt, {\it {When Energy Goes
  Missing: New Physics in $b\to s\nu\nu$ with Sterile Neutrinos}},
  \href{http://arxiv.org/abs/2309.02940}{{\tt arXiv:2309.02940}}.

\bibitem{Buras:2014fpa}
A.~J. Buras, J.~Girrbach-Noe, C.~Niehoff, and D.~M. Straub, {\it {$ B\to
  {K}^{\left(\ast \right)}\nu \overline{\nu} $ decays in the Standard Model and
  beyond}},  {\em JHEP} {\bf 02} (2015) 184,
  [\href{http://arxiv.org/abs/1409.4557}{{\tt arXiv:1409.4557}}].

\bibitem{Belle-II:2022cgf}
{\bf Belle-II} Collaboration, L.~Aggarwal et~al., {\it {Snowmass White Paper:
  Belle II physics reach and plans for the next decade and beyond}},
  \href{http://arxiv.org/abs/2207.06307}{{\tt arXiv:2207.06307}}.

\bibitem{Belle-II:2018jsg}
{\bf Belle-II} Collaboration, W.~Altmannshofer et~al., {\it {The Belle II
  Physics Book}},  {\em PTEP} {\bf 2019} (2019), no.~12 123C01,
  [\href{http://arxiv.org/abs/1808.10567}{{\tt arXiv:1808.10567}}]. [Erratum:
  PTEP 2020, 029201 (2020)].

\bibitem{Dorsner:2016wpm}
I.~Dor\v{s}ner, S.~Fajfer, A.~Greljo, J.~F. Kamenik, and N.~Ko\v{s}nik, {\it
  {Physics of leptoquarks in precision experiments and at particle colliders}},
   {\em Phys. Rept.} {\bf 641} (2016) 1--68,
  [\href{http://arxiv.org/abs/1603.04993}{{\tt arXiv:1603.04993}}].

\bibitem{Crivellin:2021lix}
A.~Crivellin, J.~F. Eguren, and J.~Virto, {\it {Next-to-leading-order QCD
  matching for \ensuremath{\Delta}F = 2 processes in scalar leptoquark
  models}},  {\em JHEP} {\bf 03} (2022) 185,
  [\href{http://arxiv.org/abs/2109.13600}{{\tt arXiv:2109.13600}}].

\bibitem{Aebischer:2020dsw}
J.~Aebischer, C.~Bobeth, A.~J. Buras, and J.~Kumar, {\it {SMEFT ATLAS of
  $\Delta$F = 2 transitions}},  {\em JHEP} {\bf 12} (2020) 187,
  [\href{http://arxiv.org/abs/2009.07276}{{\tt arXiv:2009.07276}}].

\bibitem{Isidori:2010kg}
G.~Isidori, Y.~Nir, and G.~Perez, {\it {Flavor Physics Constraints for Physics
  Beyond the Standard Model}},  {\em Ann. Rev. Nucl. Part. Sci.} {\bf 60}
  (2010) 355, [\href{http://arxiv.org/abs/1002.0900}{{\tt arXiv:1002.0900}}].

\bibitem{ATLAS:2023uox}
{\bf ATLAS} Collaboration, {\it {Search for pair production of third-generation
  leptoquarks decaying into a bottom quark and a $\tau$-lepton with the ATLAS
  detector}},  \href{http://arxiv.org/abs/2303.01294}{{\tt arXiv:2303.01294}}.

\bibitem{CMS:2018iye}
{\bf CMS} Collaboration, A.~M. Sirunyan et~al., {\it {Search for heavy
  neutrinos and third-generation leptoquarks in hadronic states of two $\tau$
  leptons and two jets in proton-proton collisions at $\sqrt{s} =$ 13 TeV}},
  {\em JHEP} {\bf 03} (2019) 170, [\href{http://arxiv.org/abs/1811.00806}{{\tt
  arXiv:1811.00806}}].

\bibitem{Azatov:2022itm}
A.~Azatov, F.~Garosi, A.~Greljo, D.~Marzocca, J.~Salko, and S.~Trifinopoulos,
  {\it {New physics in b \textrightarrow{} s\ensuremath{\mu}\ensuremath{\mu}:
  FCC-hh or a muon collider?}},  {\em JHEP} {\bf 10} (2022) 149,
  [\href{http://arxiv.org/abs/2205.13552}{{\tt arXiv:2205.13552}}].

\bibitem{Davidson:2020hkf}
S.~Davidson, {\it {Completeness and complementarity for $\mu \to e\gamma \mu
  \to e \bar e e$ and $\mu A \to eA$}},  {\em JHEP} {\bf 02} (2021) 172,
  [\href{http://arxiv.org/abs/2010.00317}{{\tt arXiv:2010.00317}}].

\bibitem{Isidori:2003ts}
G.~Isidori and R.~Unterdorfer, {\it {On the short distance constraints from
  K(L,S) ---\ensuremath{>} mu+ mu-}},  {\em JHEP} {\bf 01} (2004) 009,
  [\href{http://arxiv.org/abs/hep-ph/0311084}{{\tt hep-ph/0311084}}].

\bibitem{Dery:2021mct}
A.~Dery, M.~Ghosh, Y.~Grossman, and S.~Schacht, {\it {K \textrightarrow{}
  \ensuremath{\mu}$^{+}$\ensuremath{\mu}$^{-}$ as a clean probe of
  short-distance physics}},  {\em JHEP} {\bf 07} (2021) 103,
  [\href{http://arxiv.org/abs/2104.06427}{{\tt arXiv:2104.06427}}].

\bibitem{Brod:2022khx}
J.~Brod and E.~Stamou, {\it {Impact of indirect CP violation on
  Br(K$_{S}$\textrightarrow{}
  \ensuremath{\mu}$^{+}$\ensuremath{\mu}$^{-}$)$_{\ell=0}$}},  {\em JHEP} {\bf
  05} (2023) 155, [\href{http://arxiv.org/abs/2209.07445}{{\tt
  arXiv:2209.07445}}].

\bibitem{Hiller:2003js}
G.~Hiller and F.~Kruger, {\it {More model-independent analysis of $b \to s$
  processes}},  {\em Phys. Rev. D} {\bf 69} (2004) 074020,
  [\href{http://arxiv.org/abs/hep-ph/0310219}{{\tt hep-ph/0310219}}].

\bibitem{DAmico:2017mtc}
G.~D'Amico, M.~Nardecchia, P.~Panci, F.~Sannino, A.~Strumia, R.~Torre, and
  A.~Urbano, {\it {Flavour anomalies after the $R_{K^*}$ measurement}},  {\em
  JHEP} {\bf 09} (2017) 010, [\href{http://arxiv.org/abs/1704.05438}{{\tt
  arXiv:1704.05438}}].

\bibitem{LHCb:2022qnv}
{\bf LHCb} Collaboration, R.~Aaij et~al., {\it {Test of lepton universality in
  $b \rightarrow s \ell^+ \ell^-$ decays}},  {\em Phys. Rev. Lett.} {\bf 131}
  (2023), no.~5 051803, [\href{http://arxiv.org/abs/2212.09152}{{\tt
  arXiv:2212.09152}}].

\bibitem{LHCb:2022vje}
{\bf LHCb} Collaboration, .~\textrightarrow{} K~+ +~\ensuremath{-} et~al., {\it
  {Measurement of lepton universality parameters in $B^+\to K^+\ell^+\ell^-$
  and $B^0\to K^{*0}\ell^+\ell^-$ decays}},  {\em Phys. Rev. D} {\bf 108}
  (2023), no.~3 032002, [\href{http://arxiv.org/abs/2212.09153}{{\tt
  arXiv:2212.09153}}].

\bibitem{LHCb:2018roe}
{\bf LHCb} Collaboration, R.~Aaij et~al., {\it {Physics case for an LHCb
  Upgrade II - Opportunities in flavour physics, and beyond, in the HL-LHC
  era}},  \href{http://arxiv.org/abs/1808.08865}{{\tt arXiv:1808.08865}}.

\bibitem{Bordone:2016gaq}
M.~Bordone, G.~Isidori, and A.~Pattori, {\it {On the Standard Model predictions
  for $R_K$ and $R_{K^*}$}},  {\em Eur. Phys. J. C} {\bf 76} (2016), no.~8 440,
  [\href{http://arxiv.org/abs/1605.07633}{{\tt arXiv:1605.07633}}].

\bibitem{Isidori:2022bzw}
G.~Isidori, D.~Lancierini, S.~Nabeebaccus, and R.~Zwicky, {\it {QED in $
  \overline{B} $\textrightarrow{}$ \overline{K}
  $\ensuremath{\ell}$^{+}$\ensuremath{\ell}$^{-}$ LFU ratios: theory versus
  experiment, a Monte Carlo study}},  {\em JHEP} {\bf 10} (2022) 146,
  [\href{http://arxiv.org/abs/2205.08635}{{\tt arXiv:2205.08635}}].

\bibitem{Crivellin:2018qmi}
A.~Crivellin, M.~Hoferichter, and P.~Schmidt-Wellenburg, {\it {Combined
  explanations of $(g-2)_{\mu,e}$ and implications for a large muon EDM}},
  {\em Phys. Rev. D} {\bf 98} (2018), no.~11 113002,
  [\href{http://arxiv.org/abs/1807.11484}{{\tt arXiv:1807.11484}}].

\bibitem{Isidori:2021gqe}
G.~Isidori, J.~Pag\`es, and F.~Wilsch, {\it {Flavour alignment of New Physics
  in light of the (g \ensuremath{-} 2)$_{\mu}$ anomaly}},  {\em JHEP} {\bf 03}
  (2022) 011, [\href{http://arxiv.org/abs/2111.13724}{{\tt arXiv:2111.13724}}].

\bibitem{MEG:2016leq}
{\bf MEG} Collaboration, A.~M. Baldini et~al., {\it {Search for the lepton
  flavour violating decay $\mu ^+ \rightarrow \mathrm {e}^+ \gamma $ with the
  full dataset of the MEG experiment}},  {\em Eur. Phys. J. C} {\bf 76} (2016),
  no.~8 434, [\href{http://arxiv.org/abs/1605.05081}{{\tt arXiv:1605.05081}}].

\bibitem{MEGII:2018kmf}
{\bf MEG II} Collaboration, A.~M. Baldini et~al., {\it {The design of the MEG
  II experiment}},  {\em Eur. Phys. J. C} {\bf 78} (2018), no.~5 380,
  [\href{http://arxiv.org/abs/1801.04688}{{\tt arXiv:1801.04688}}].

\bibitem{EuropeanStrategyforParticlePhysicsPreparatoryGroup:2019qin}
R.~K. Ellis et~al., {\it {Physics Briefing Book}: {Input for the European
  Strategy for Particle Physics Update 2020}},
  \href{http://arxiv.org/abs/1910.11775}{{\tt arXiv:1910.11775}}.

\bibitem{Lee:2018pag}
L.~Lee, C.~Ohm, A.~Soffer, and T.-T. Yu, {\it {Collider Searches for Long-Lived
  Particles Beyond the Standard Model}},  {\em Prog. Part. Nucl. Phys.} {\bf
  106} (2019) 210--255, [\href{http://arxiv.org/abs/1810.12602}{{\tt
  arXiv:1810.12602}}]. [Erratum: Prog.Part.Nucl.Phys. 122, 103912 (2022)].

\bibitem{Calibbi:2021fld}
L.~Calibbi, F.~D'Eramo, S.~Junius, L.~Lopez-Honorez, and A.~Mariotti, {\it
  {Displaced new physics at colliders and the early universe before its first
  second}},  {\em JHEP} {\bf 05} (2021) 234,
  [\href{http://arxiv.org/abs/2102.06221}{{\tt arXiv:2102.06221}}].

\bibitem{CMS:2018oaj}
{\bf CMS} Collaboration, A.~M. Sirunyan et~al., {\it {Search for leptoquarks
  coupled to third-generation quarks in proton-proton collisions at $\sqrt{s}=$
  13 TeV}},  {\em Phys. Rev. Lett.} {\bf 121} (2018), no.~24 241802,
  [\href{http://arxiv.org/abs/1809.05558}{{\tt arXiv:1809.05558}}].

\bibitem{ATLAS:2020xyo}
{\bf ATLAS} Collaboration, G.~Aad et~al., {\it {Search for long-lived, massive
  particles in events with a displaced vertex and a muon with large impact
  parameter in $pp$ collisions at $\sqrt{s} = 13$ TeV with the ATLAS
  detector}},  {\em Phys. Rev. D} {\bf 102} (2020), no.~3 032006,
  [\href{http://arxiv.org/abs/2003.11956}{{\tt arXiv:2003.11956}}].

\bibitem{ATLAS:2022pib}
{\bf ATLAS} Collaboration, G.~Aad et~al., {\it {Search for heavy, long-lived,
  charged particles with large ionisation energy loss in $pp$ collisions at
  $\sqrt{s} = 13~\text{TeV}$ using the ATLAS experiment and the full Run 2
  dataset}},  {\em JHEP} {\bf 2306} (2023) 158,
  [\href{http://arxiv.org/abs/2205.06013}{{\tt arXiv:2205.06013}}].

\bibitem{Gross:2018zha}
C.~Gross, A.~Mitridate, M.~Redi, J.~Smirnov, and A.~Strumia, {\it {Cosmological
  Abundance of Colored Relics}},  {\em Phys. Rev. D} {\bf 99} (2019), no.~1
  016024, [\href{http://arxiv.org/abs/1811.08418}{{\tt arXiv:1811.08418}}].

\bibitem{Kawasaki:2017bqm}
M.~Kawasaki, K.~Kohri, T.~Moroi, and Y.~Takaesu, {\it {Revisiting Big-Bang
  Nucleosynthesis Constraints on Long-Lived Decaying Particles}},  {\em Phys.
  Rev. D} {\bf 97} (2018), no.~2 023502,
  [\href{http://arxiv.org/abs/1709.01211}{{\tt arXiv:1709.01211}}].

\bibitem{FCC:2018vvp}
{\bf FCC} Collaboration, A.~Abada et~al., {\it {FCC-hh: The Hadron Collider}:
  {Future Circular Collider Conceptual Design Report Volume 3}},  {\em Eur.
  Phys. J. ST} {\bf 228} (2019), no.~4 755--1107.

\bibitem{MoEDAL-MAPP:2022kyr}
{\bf MoEDAL-MAPP} Collaboration, B.~Acharya et~al., {\it {MoEDAL-MAPP, an LHC
  Dedicated Detector Search Facility}},  in {\em {Snowmass 2021}}, 9, 2022.
\newblock \href{http://arxiv.org/abs/2209.03988}{{\tt arXiv:2209.03988}}.

\bibitem{Esteban:2020cvm}
I.~Esteban, M.~C. Gonzalez-Garcia, M.~Maltoni, T.~Schwetz, and A.~Zhou, {\it
  {The fate of hints: updated global analysis of three-flavor neutrino
  oscillations}},  {\em JHEP} {\bf 09} (2020) 178,
  [\href{http://arxiv.org/abs/2007.14792}{{\tt arXiv:2007.14792}}].

\bibitem{Minkowski:1977sc}
P.~Minkowski, {\it {$\mu \to e\gamma$ at a Rate of One Out of $10^{9}$ Muon
  Decays?}},  {\em Phys. Lett. B} {\bf 67} (1977) 421--428.

\bibitem{Yanagida:1980xy}
T.~Yanagida, {\it {Horizontal Symmetry and Masses of Neutrinos}},  {\em Prog.
  Theor. Phys.} {\bf 64} (1980) 1103.

\bibitem{Gell-Mann:1979vob}
M.~Gell-Mann, P.~Ramond, and R.~Slansky, {\it {Complex Spinors and Unified
  Theories}},  {\em Conf. Proc. C} {\bf 790927} (1979) 315--321,
  [\href{http://arxiv.org/abs/1306.4669}{{\tt arXiv:1306.4669}}].

\bibitem{Pati:1974yy}
J.~C. Pati and A.~Salam, {\it {Lepton Number as the Fourth Color}},  {\em Phys.
  Rev. D} {\bf 10} (1974) 275--289. [Erratum: Phys.Rev.D 11, 703--703 (1975)].

\bibitem{FileviezPerez:2013zmv}
P.~Fileviez~Perez and M.~B. Wise, {\it {Low Scale Quark-Lepton Unification}},
  {\em Phys. Rev. D} {\bf 88} (2013) 057703,
  [\href{http://arxiv.org/abs/1307.6213}{{\tt arXiv:1307.6213}}].

\bibitem{FileviezPerez:2023umx}
P.~Fileviez~Perez, C.~Murgui, S.~Patrone, A.~Testa, and M.~B. Wise, {\it
  {Finite Naturalness and Quark-Lepton Unification}},
  \href{http://arxiv.org/abs/2308.07367}{{\tt arXiv:2308.07367}}.

\bibitem{Maggiore:2019uih}
M.~Maggiore et~al., {\it {Science Case for the Einstein Telescope}},  {\em
  JCAP} {\bf 03} (2020) 050, [\href{http://arxiv.org/abs/1912.02622}{{\tt
  arXiv:1912.02622}}].

\bibitem{Evans:2021gyd}
M.~Evans et~al., {\it {A Horizon Study for Cosmic Explorer: Science,
  Observatories, and Community}},  \href{http://arxiv.org/abs/2109.09882}{{\tt
  arXiv:2109.09882}}.

\bibitem{Martin:2019lqd}
S.~P. Martin and D.~G. Robertson, {\it {Standard model parameters in the
  tadpole-free pure $\overline{\rm{MS}}$ scheme}},  {\em Phys. Rev. D} {\bf
  100} (2019), no.~7 073004, [\href{http://arxiv.org/abs/1907.02500}{{\tt
  arXiv:1907.02500}}].

\bibitem{Thomsen:2021ncy}
A.~E. Thomsen, {\it {Introducing RGBeta: a Mathematica package for the
  evaluation of renormalization group $ \beta $-functions}},  {\em Eur. Phys.
  J. C} {\bf 81} (2021), no.~5 408,
  [\href{http://arxiv.org/abs/2101.08265}{{\tt arXiv:2101.08265}}].

\bibitem{Workman:2022ynf}
{\bf Particle Data Group} Collaboration, R.~L. Workman and Others, {\it {Review
  of Particle Physics}},  {\em PTEP} {\bf 2022} (2022) 083C01.

\end{thebibliography}\endgroup

\end{document}